\documentclass[twocolumn]{aastex61}
\usepackage{graphicx}
\usepackage{subfigure}
\usepackage{multirow}
\usepackage{comment}
\usepackage{natbib}
\usepackage{hyperref}
\usepackage{mathtools}
\usepackage{mathrsfs}
\usepackage{fontenc}
\usepackage{color}
\usepackage{url}
\usepackage{hyperref}
\usepackage{gensymb}
\usepackage{pifont}
\usepackage{dcolumn}
\newcolumntype{d}[1]{D{.}{.}{#1}}

\newcommand{\tnm}{\tablenotemark}
\newcommand{\mcl}{\multicolumn}
\newcommand{\ph}{\phantom}

\def\hst{{\it HST}}
\def\herschel{{\it Herschel}}
\def\pacs{{PACS}}
\def\spire{{SPIRE}}
\def\galfit{{\tt GALFIT}}
\def\tmass{{2MASS}}
\def\wise{{\it WISE}}
\def\w{{\it W}}
\def\spitzer{{\it Spitzer}}

\newcommand{\kms}{${\rm km~ s^{-1}}$}

\newcommand{\hi}{H{\sevenrm\,I}}
\newcommand{\OIII}{[O{\sevenrm\,III}]}
\newcommand{\OIIIb}{[O{\sevenrm\,III}]\,$\lambda$5007}

\newcommand{\loiii}{$L_{\text{[O \textrm{\tiny III}]}}$}
\newcommand{\umin}{$U_\mathrm{min}$}
 
 \font\sevenrm=cmr7 scaled 1000

\newcommand{\mzr}{$M_*$--$Z$}
\newcommand{\gzr}{$\delta_\mathrm{GDR}$--$Z$}
\newcommand{\gdr}{$\delta_\mathrm{GDR}$}
\newcommand{\cohii}{CO-to-$\mathrm{H}_2$}
\newcommand{\PaperI}{\cite{Shangguan2018ApJ}}

\begin{document}

\title{Testing the Evolutionary Link between Type 1 and Type 2 Quasars with 
Measurements of the Interstellar Medium}

\author{Jinyi Shangguan}
\affil{Kavli Institute for Astronomy and Astrophysics, Peking University,
Beijing 100871, China}
\affil{Department of Astronomy, School of Physics, Peking University,
Beijing 100871, China}

\author{Luis C. Ho}
\affil{Kavli Institute for Astronomy and Astrophysics, Peking University,
Beijing 100871, China}
\affil{Department of Astronomy, School of Physics, Peking University,
Beijing 100871, China}

\begin{abstract}
In a popular scenario for the coevolution of massive black holes and galaxies,
major mergers of gas-rich galaxies fuel vigorous star formation and obscured
(type~2) quasar activity until energy feedback from the active galactic nucleus
clears away the gas and dust to reveal an unobscured (type~1) quasar.
Under this scenario, the precursor type~2 quasars should be more gas-rich than
their type~1 counterparts, and both types of quasars are expected to be
gas-deficient relative to normal, star-forming galaxies of similar stellar
mass.  We test this evolutionary hypothesis by investigating the infrared
($\sim 1-500$ \micron) spectral energy distribution of 86 optically selected
$z < 0.5$ type~2 quasars, matched in redshift and \OIII\ luminosity to a
comparison sample of type~1 quasars.  Contrary to expectations, the gas
content of the host galaxies of type~2 quasars is nearly indistinguishable
from that of type~1 quasar hosts, and neither type exhibits the predicted
deficit in gas relative to normal galaxies.  The gas mass fraction of quasar
hosts appears unaffected by the bolometric luminosity of the active nucleus, 
although their interstellar radiation field is preferentially higher than that 
of normal galaxies, potentially implicating active galactic nucleus heating of 
the large-scale galactic dust.
\end{abstract}

\keywords{galaxies: active --- galaxies: ISM --- galaxies: nuclei ---
galaxies: Seyfert --- (galaxies:) quasars: general --- infrared: ISM}

\section{Introduction}

Mergers of gas-rich galaxies, which lead to loss of angular momentum of the 
gas and gas inflows, are often invoked as the mechanism to trigger the most
powerful active galactic nuclei (AGNs) and quasars \citep{Heckman1986ApJ,
Sanders1988ApJ,Jogee2006LNP}, which are accompanied or immediately 
preceded by a powerful starburst \citep{Hopkins2006ApJS}.  The complex, 
chaotic distribution of the dusty interstellar medium (ISM) during the early 
phases of the merger process results in highly obscured black hole growth.  
The most luminous AGNs activated during this period are likely ``type~2 quasars" 
(QSO2s).  Once the black hole reaches a sufficiently large mass, the AGN 
can clear the gas and dust and reveal itself as a type~1 quasar (QSO1;
\citealt{Hopkins2008ApJS}).  Quasar-mode AGN feedback injects significant 
energy into the host galaxy and quenches galaxy-wide star formation, explosively 
removing the cold ISM from the host galaxy \citep{Silk1998AA,Fabian2012ARAA}.  
This scenario also provides a natural explanation for the coevolution of the black 
hole and its host galaxy \citep{KH13}, which is implicated by the observed tight 
correlation between the mass of the black hole and the mass and velocity 
dispersion of the stellar bulge \citep{Magorrian1998AJ,Gebhardt2000ApJ,
Ferrarese2000ApJ}.  Alternatively, one might explain the observed differences 
between QSO1s and QSO2s simply through viewing angle-dependent 
obscuration by a small-scale dusty torus, following the traditional AGN 
``unified model'' \citep{Antonucci1993ARAA} originally proposed for lower 
luminosity Seyfert galaxies.

There have been many attempts to observationally test the merger-driven model
for AGN evolution, but the results are still controversial.  On the one hand,
the host galaxies of QSO1s and QSO2s appear to have a number of
differences, including their star formation rates (e.g., \citealt{Kim2006ApJ,
Chen2015ApJ}), ionized gas velocity fields \citep{Greene2011ApJ}, radio
continuum properties \citep{Lal2010AJ}, and local environment
\citep{Villarroel2014NatPh}.  The apparent differences between the host
galaxies of the two quasar types suggest that they are intrinsically
different.  If so, the two quasar types are linked by an evolutionary
connection instead of purely by viewing-angle orientation.  On the other hand,
unequivocal evidence for the role of mergers or interactions in triggering AGN
activity remains elusive.  While some studies find that the incidence of AGNs 
increases in close galaxy pairs (e.g., \citealt{Silverman2011ApJ}), 
especially those with  decreasing physical separation \citep{Ellison2011MNRAS}, 
others fail to find a clear enhancement of
merger features in {\it Hubble Space Telescope} (\hst) images of AGN and quasar
host galaxies (e.g., \citealt{Grogin2005ApJ,Cisternas2011ApJ,Kocevski2012ApJ,
Bohm2013AA,Villforth2014MNRAS,Mechtley2016ApJ,Villforth2017MNRAS,Zhao2019}).  Such
morphological studies, however, are subject to uncertainties due to image
depth (\citealt{Bennert2008ApJ,Hong2015ApJ}), image contrast with the bright
nucleus, time lag between galaxy coalescence and the onset of AGN activity
(\citealt{Bohm2013AA,Villforth2017MNRAS}), and possible biases stemming from
X-ray sample selection (e.g., \citealt{Kocevski2015ApJ,Shangguan2016ApJ}).

The above studies largely focus on the stars of the host galaxy.  The ISM
component of the host provides complementary insights.  The evolutionary
scenario naturally predicts that QSO1 host galaxies should be more
gas-deficient than QSO2 hosts as a result of efficient blow-out of cold gas by
AGN feedback during the QSO1 phase.  If AGN activity turns on with a
substantial time delay (e.g., $\gtrsim 1\,\mathrm{Gyr}$) after the onset of
starburst activity,
we also expect quasar host galaxies (of either type) to have systematically
reduced gas content compared to normal galaxies of similar stellar mass, even
for star formation rates of relatively modest intensity (e.g.,
$\sim 10\,M_\odot\,\mathrm{yr^{-1}}$).  While observations of the CO molecule
can probe molecular gas in AGNs over a wide range of redshifts and luminosities,
from nearby lower luminosity sources \citep{Scoville2003ApJ,Evans2006AJ,
Bertram2007AA,Husemann2017arXiv} to powerful quasars out to $z\gtrsim6$ (e.g.,
\citealt{Wang2013ApJ,Cicone2014AA,Walter2004ApJ,Wang2016ApJ}), they are still
quite time-consuming to make and plagued by uncertainty from the \cohii\
conversion factor \citep{Bolatto2013ARAA}.  In the mean time, \hi\ observations
currently can hardly extend beyond $z \approx 0.2$, where most quasars lie.
Thanks to the unprecedented sensitivity and spatial resolution of the {\it
Herschel Space Observatory}\ \citep{Pilbratt2010AA}, we can probe the cold ISM
accurately and efficiently with dust emission in the far-infrared (FIR).

\PaperI\ recently developed a new method to measure dust masses from 
detailed modeling of the IR spectral energy distribution (SED) of quasars, 
from which robust total (atomic and molecular) gas masses can 
be derived using gas-to-dust ratios estimated from the metallicity and the
mass-metallicity relation.  They applied their technique to study the ISM
content of the sample of 87 $z<0.5$ QSO1s from the Palomar-Green (PG;
\citealt{Schmidt1983ApJ}) survey.  Here we focus our attention on QSO2s,
choosing a sample well-matched to the PG QSO1s, with the intent of
investigating the possible evolutionary connection between these two types 
of luminous AGNs.  We measure the dust masses of QSO2s from their 
photometric SEDs and estimate the total gas masses based on the dust content, 
closely following the methodology developed for our previous study of QSO1s 
\citep{Shangguan2018ApJ}.  
We find that the hosts of QSO1s and QSO2s have surprisingly similar 
ISM properties.  Both quasar types also turn out to have dust and gas 
fractions comparable to those of normal, star-forming galaxies of similar 
stellar mass.  This result is in apparent conflict with the most basic expectation 
of the merger-driven evolutionary scenario for massive galaxies.

Luminous AGNs have diverse SEDs that may evolve with time (e.g., 
\citealt{Haas2003AA}).  While it is beyond the scope of this work to sample
the full range of luminous AGN properties, our study is designed to detect 
differences in gas mass between the two types of quasars, differences that are
inevitable so long as quasar-mode feedback plays a critical in transforming 
QSO2s to QSO1s.  Our experiment probes the total cold gas content, not just the 
star-forming molecular medium as envisioned in the original evolutionary model 
of \cite{Sanders1988ApJ}.  Nevertheless, for AGN feedback to substantially 
influence galaxy evolution, it must affect the bulk of the cold ISM, which, in 
any event, is not distinguished into molecular or atomic phase in numerical 
simulations (e.g., \citealt{Genel2014MNRAS,Lagos2015MNRAS}).  AGN outflows are 
widely regarded as multiphased (e.g., \citealt{Harrison2018NatAs} and 
references therein).

The paper is organized as follows.  Section \ref{sec:data} describes the quasar
samples and data reduction to construct the near-IR (NIR) to FIR SED.  The
results of our measurements, including the stellar mass, interstellar
radiation field intensity, and the dust and gas masses, are presented
in Section \ref{sec:results}.  In Section \ref{sec:discuss}, we investigate
whether the differences between the SEDs of QSO1s and QSO2s can be explained
by dust extinction and discuss the implications of our results in terms of the
evolutionary scenario for quasars.  This work adopts the following parameters
for a $\Lambda$CDM cosmology: $\Omega_m = 0.308$, $\Omega_\Lambda = 0.692$, and
$H_{0}=67.8$ km s$^{-1}$ Mpc$^{-1}$ \citep{Planck2015arXiv}.

\section{Sample and Data Reduction}
\label{sec:data}

\subsection{Quasar Samples}
\label{ssec:sample}

\begin{figure}[htbp]
\begin{center}
\includegraphics[height=0.35\textheight]{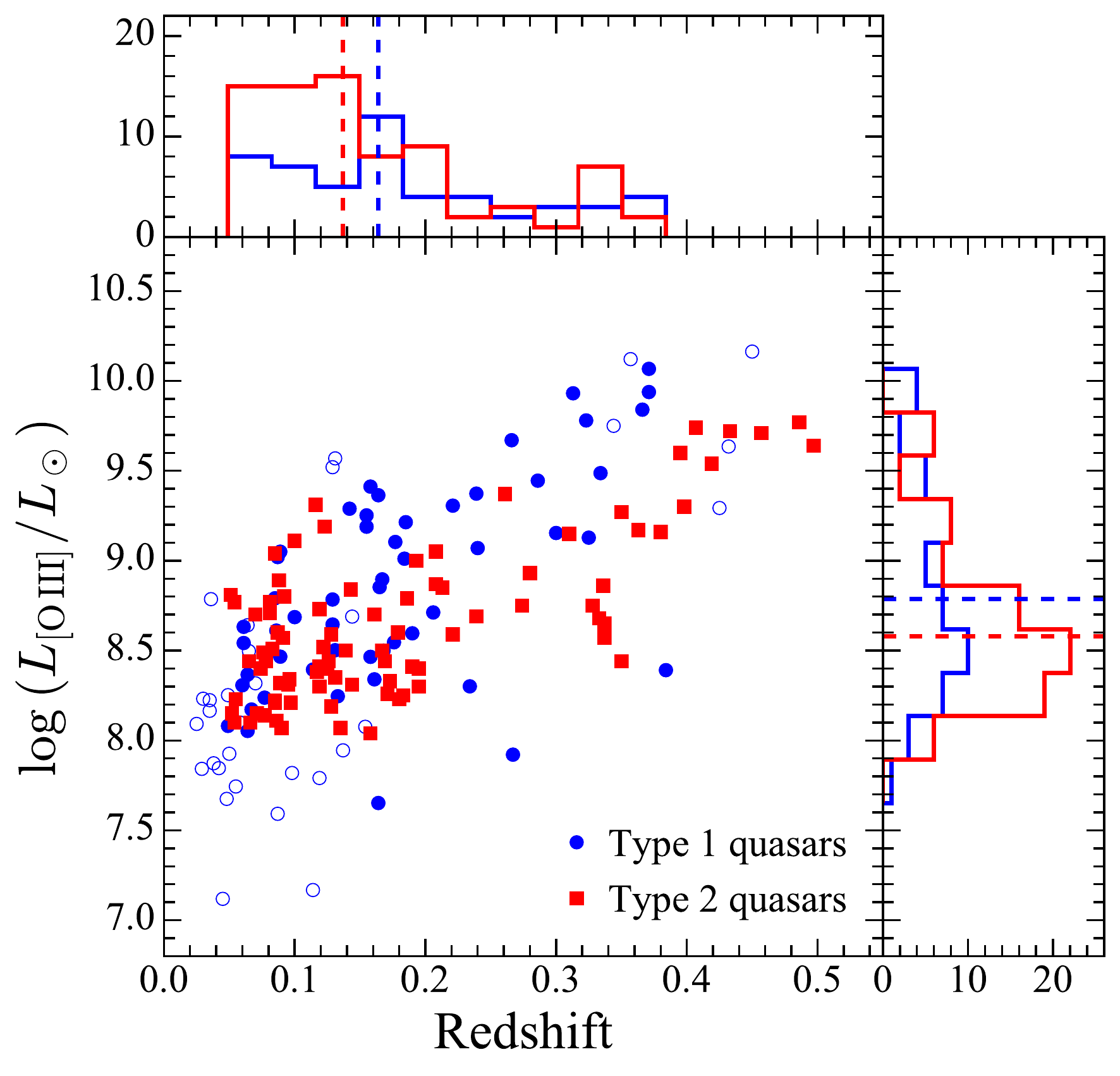}
\caption{The distribution of redshift and \OIII\ $\lambda 5007$ luminosity of 
the samples of type~1 and type~2 quasars. The two samples are matched in terms 
of these two quantities.  The filled blue circles are the type~1 quasars with 
host galaxy stellar mass measurements.  We will use these objects to compare 
with the type~2 quasars, while the rest of the objects are in empty circles.  
The side panels show the distributions of the filled symbols of the type~1 and 
type~2 quasars.  The dashed lines are the median values of the corresponding 
filled symbols.  The type~1 quasars show on average $\sim$0.2 dex 
higher \OIII\ luminosity and $\sim$0.03 dex higher redshift, but these 
differences are small and unlikely to affect our results.}
\label{fig:sample}
\end{center}
\end{figure}

Given our main goal of testing the hypothesis that QSO2s are the progenitors
of QSO1s, we select closely matched samples of the two quasar types.  Our
reference sample of QSO1s are the 87 low-redshift ($z < 0.5$)
optical/UV-selected PG quasars of \cite{Boroson1992ApJS}, whose ISM
properties are described in the companion paper by \PaperI.  We choose
87 QSO2s (Table \ref{tab:results}) derived from the Sloan Digital Sky Survey
(SDSS; \citealt{Reyes2008AJ}), randomly selected to match the PG QSO1s in both
redshift and \OIII\ $\lambda 5007$ luminosity (Figure \ref{fig:sample}).  Following the earlier
work of \cite{Zakamska2003AJ}, \cite{Reyes2008AJ} identified QSO2s as
extragalactic sources that have (1) optical diagnostic emission-line intensity
ratios consistent with AGN excitation, (2) sufficiently narrow (FWHM $< 2000$
\kms) permitted lines that make them likely candidates for type~2 sources,
and (3) \OIII\ luminosities larger than $2\times 10^{8}\ L_{\odot}$, which,
when translated to equivalent $B$-band absolute magnitudes, qualify them
as quasars by the historical criterion of $M_B < -23$ mag \citep{Schmidt1983ApJ}.
If an evolutionary link exists between the two types of quasars, we expect
the gas masses of the QSO2 hosts to be systematically higher than those
of the QSO1 hosts, for a given host galaxy stellar mass.  Based on an X-ray
study of QSO2s from our parent sample \citep{Jia2013ApJ}, we expect
that more than half of our QSO2s are Compton-thick, and most of the
rest of the objects should be considerably obscured.

For direct comparison with the QSO2s, we mainly focus on the subset of 55
QSO1s having stellar masses estimated from decomposition of high-resolution 
images \citep{Zhang2016ApJ}.  Due to this additional constraint, the 
median properties of the two samples are slightly different, on average by 
$\sim$0.2 dex in \OIII\ luminosity and $\sim$0.03 dex in redshift, but the 
mismatch is small compared to the sample range and will not affect our 
conclusions.

Galaxy mergers play a role in triggering nuclear activity in the most 
powerful unobscured \citep{Bahcall1997ApJ,McLure1999MNRAS,Dunlop2003MNRAS,
Letawe2010MNRAS,Hong2015ApJ} and obscured \citep{Bessiere2012MNRAS} quasars.  
Internal processes may be more relevant for activating nuclei
of lower luminosity \citep{Hopkins2009ApJ,Treister2012ApJ,
Villforth2017MNRAS}.  Limitations in sensitivity and resolution make it very 
challenging to confirm whether any given host galaxy has experienced a merger, 
as tidal features can be faint and hard to detect,
especially in the presence of an overpowering bright nucleus.  Deep,
ground-based imaging studies \citep{Letawe2010MNRAS,Hong2015ApJ} reveal a high
incidence of tidal features and other morphological signatures suggestive of
mergers in quasar host galaxies.  We therefore work under the assumption that
most of the quasars in our sample are triggered by mergers.

Part of our analysis will be restricted solely to the subset of QSO1s and QSO2s 
that are clearly hosted by galaxy mergers (Section 3.5).  The information for 
QSO1s is mainly based on archival data \citep{Kim2008ApJS,Kim2017ApJS} from 
\hst\ and pointed observations from our own ongoing \hst\ project (Y. Zhao et 
al., in preparation), supplemented by ground-based observations 
\citep{Hong2015ApJ}.  One-third (29/87) of the QSO2s were observed by our own 
{\it HST}\ project (Zhao et al.  2019), and the rest of the sample were 
examined with SDSS images, which, unfortunately, can hardly reveal galaxy 
merger features beyond $z \approx 0.15$.  In total there are 11 QSO1s and 15 
QSO2s with merger features at $z < 0.15$.

\subsection{2MASS and WISE}

Emission from evolved stars of the host galaxy dominates the
\tmass\ \citep{Skrutskie2006AJ} $J$ (1.235 $\mu$m), $H$ (1.662 $\mu$m), and
$K_s$ (2.159 $\mu$m) bands \citep{Cohen2003AJ}.  We perform aperture 
photometry for the \tmass\ data to obtain source flux
densities or their respective upper limits.  We collect the \tmass\ images
from the NASA/IPAC Infrared Science Archive (IRSA)\footnote{\url{
irsa.ipac.caltech.edu/frontpage/}} by matching each source with a
$5\arcsec$ radius, and the measurements are conducted using the Python package
{\tt photutils}\footnote{\url{http://photutils.readthedocs.io/en/stable/}}.
We first fit and subtract the background using a third-order two-dimensional
polynomial function to remove possible large-scale gradients, mainly due to
``airglow'' emission \citep{Jarrett2000AJ}.  To measure the integrated flux of
the source, we use the default aperture radius of 7\arcsec\
\citep{Jarrett2000AJ} with the sky annulus set to a radius of 25\arcsec\ to
35\arcsec.  For the nearest ($z\lesssim0.1$) quasars with more extended host
galaxies, we use larger aperture radii but the same sky annulus.  However,
larger aperture includes more noise.  We, therefore, carefully choose the
aperture radius (10\arcsec, 15\arcsec, 20\arcsec) to ensure that the flux
of these extended sources is not below 3 times the uncertainty\footnote{In
contrast, we uniformly use a 20\arcsec\ aperture for the extended hosts of PG
quasars \citep{Shangguan2018ApJ} because they are all detected.}.  The
selected apertures, which correspond to at least 12 kpc in physical size for
our sample, are large enough to cover the entire galaxy.  To determine the
uncertainty, we perform 500 random aperture measurements of the sky, in
exactly the same way as the quasar, with all sources masked, and calculate the
standard deviation.  We do not apply any aperture correction, which is found
very small\footnote{\url{www.astro.caltech.edu/\~jmc/2mass/v3/images/}}.  The
measurements of eight targets are affected by projected close companions (see
Table \ref{tab:nirdata}).  We first use \galfit\ \citep{Peng2002AJ,Peng2010AJ}
to fit and remove the companions from the images.  The point-spread function
(PSF) of each image is derived from the stars in the field using {\tt DAOPHOT}
in IRAF\footnote{IRAF is distributed by the National Optical Astronomy
Observatories, which are operated by the Association of Universities for
Research in Astronomy, Inc., under cooperative agreement with the National
Science Foundation.} \citep{Tody1986SPIE}.  The residual (companion-subtracted)
images are then measured using the same method described above.  The
measurements and the aperture radii are listed in Table \ref{tab:nirdata}.

In order to obtain accurate measurements that avoid the influence of projected
companions, we also perform our own aperture photometry on the \wise\ images,
following the method applied to the \tmass\ data.  We similarly collect \wise\
\citep{Wright2010AJ,Jarrett2011ApJ} \w1 (3.353 \micron), \w2 (4.603 \micron),
\w3 (11.561 \micron), and \w4 (22.088 \micron) data from IRSA.  We perform
aperture photometry using ``standard'' aperture radii \citep{Cutri2012wise} of 
8\farcs25 for the \w1, \w2, and \w3 bands and 16\farcs5 for the \w4 band,
adopting a sky annulus of 50\arcsec--70\arcsec.  We use the curves of growth
of coadded PSFs \citep{Cutri2012wise} of the four \wise\ bands to calculate
aperture correction factors.  The uncertainty is also estimated by making 500
random measurements throughout the sky region.  We carefully check the images
and find 14 objects with close companions that show a flux density drop of
more than 5\% after we remove the companion(s) with \galfit\ (Table
\ref{tab:nirdata}).  Due to the differences in wavelength and resolution, the
projected companions in the \wise\ images are not necessarily the same as 
those in the \tmass\ images.  Although the resolution of the \wise\ images is 
low (6\farcs1, 6\farcs8, 7\farcs4, and 12\farcs0 for \w1, \w2, \w3, and
\w4, respectively), some objects with $z \lesssim 0.1$ tend to be
marginally resolved in the first three \wise\ bands. We identify five objects
whose flux densities increase by $\sim 20\%-30\%$ when we measure with
$20\arcsec$ aperture radius (Table \ref{tab:nirdata}).  Due to the low
resolution, a larger aperture is more likely to be contaminated by blended
faint sources.  We decide to keep the measurements with the 8\farcs25 aperture
radius, but enlarge the uncertainty of the \w1--\w3 results to 30\% of the flux
densities.  For the rest of the objects, mainly concerning those with
$z\lesssim 0.1$, the $<20\%$ flux decrease due to the small aperture will not
affect our SED fitting.  Both \tmass\ \citep{Jarrett2003AJ} and \wise\
\citep{Jarrett2011ApJ} have a calibration uncertainty of 3\%; this is not
included in Table \ref{tab:nirdata}.  We note that our main statistical
results are not affected by whether or not we include the objects with
possibly larger uncertainty.

\subsection{Herschel}

The majority (84/87) of the QSO2 sample was observed by our own dedicated
\herschel\ program (OT2\_lho\_2; PI: L. Ho) with both the Photodetector Array
Camera and Spectrometer (\pacs, \citealt{Poglitsch2010AA}) and the Spectral
and Photometric Imaging Receiver (\spire, \citealt{Griffin2010AA}).  The
observation for one of the objects failed because of an error in the input
coordinates.  The remaining three sources were already observed by \spire\ by
other programs (SDSS J0843$+$2944, KPOT\_gsmith01\_1, PI: T. Mueller; SDSS
J1034$+$6001, SDP\_soliver\_3, PI: S. Oliver; SDSS J1218$+$0222,
KPOT\_jdavie01\_1, PI: J. Davies).  Thus, there are 86 objects with \herschel\
measurements, forming the QSO2 sample considered in this study.

We quote monochromatic flux densities at 70, 100, and 160 \micron\ for \pacs,
and at 250, 350, and 500 \micron\ for \spire\ (Table \ref{tab:herschel}).  The
objects possibly affected by confusion from close companions, likely having 
larger uncertainties, are marked in the table.  Our results,
however, are not affected by whether or not these objects are included in the
analysis.  The standard pipeline assumes a constant $\nu f_\nu$ spectrum.  We
provide $3\,\sigma$ upper limits for non-detections.  The calibration
uncertainties for \pacs\ and \spire\ photometry are both $5\%$, which are not
included in the uncertainties quoted in Table \ref{tab:herschel}.  We do not
apply a color correction but do consider the instrument spectral response
functions in the SED modeling.

\subsubsection{PACS}
\label{ssec:pacs}
The QSO2 sample, like the PG QSO1s, is observed by PACS in
mini-scan mode, with scan angles 70\degree\ and 110\degree\ and a
scanning speed of 20\arcsec\ per second.  PACS simultaneously scans 
the sources at short (70 or 100 \micron) and long (160 \micron) wavelengths 
over a field-of-view of $1\farcm75 \times 3\farcm5$.  The integration time 
for each scan angle is 180 s.

The data are processed within version 15.0.1 of the Herschel Interactive
Processing Environment (HIPE; \citealt{Ott2010ASPC}), using the latest
calibration files (calibration tree version 78).  We use the script ``Scanmap
Pointsources PhotProject'' provided by HIPE to reduce the timeline
(\texttt{level1}) data into science images.  In order to generate a proper
mask for the high-pass filtering process to remove the ``$1/f$ noise'', a
S/N-based mask is first generated.  All the pixels above the $3\,\sigma$
threshold are masked.  Then, a circular mask with $\mathtt{radius}=25\arcsec$
is added at the nominal position of the target.  The scan maps with different
scan directions are drizzle-combined with the {\tt photProject} function,
using the default pixel fraction ($\mathtt{pixfrac}=1.0$) and reduced output
pixel size of 1\farcs1, 1\farcs4, and 2\farcs1 for the three bands,
respectively.  A smaller pixel fraction can, in principle, reduce the
covariance noise, but we find that the noise does not significantly change
when we set $\mathtt{pixfrac}=0.6$.  The key parameters, described above,
follow those used by Balog et al. (2014; their Section 4.1).

The flux density of the compact sources can be measured from aperture
photometry with the aperture sizes and annulus radii for background
subtraction following the recommendations of the \herschel\ Webinar
``Photometry Guidelines for PACS data" by
Paladini\footnote{\url{https://nhscsci.ipac.caltech.edu/workshop/Workshop\_Oct2014/Photometry/PACS/PACS\_phot\_Oct2014\_photometry.pdf}}.
The aperture radii for bright sources are $12\arcsec$, $12\arcsec$, and
$22\arcsec$ for 70, 100, and 160 \micron\ bands, respectively, and the
corresponding values for faint sources are $5\farcs5$, $5\farcs6$, and
$10\farcs5$.  The inner and outer radius of the sky annulus are 35\arcsec\
and 45\arcsec, out to which the sky measurements are affected by the
PSF wings less than $0.1\%$ \citep{Balog2014ExA}.  Aperture correction
is always necessary because the PSFs of PACS maps are very extended
(see Table 2 of \citealt{Balog2014ExA}).

As our quasars span redshifts $\sim 0.05-0.5$ with some displaying
extended tidal features, it is important to carefully assess whether the
aperture is large enough to enclose all the extended emission belonging to
the targets, especially for the low-$z$ objects.  In order to determine the
aperture radius for PG QSO1s in a similar redshift range, \PaperI\ find that
objects fainter than 200 mJy at 100 \micron\ can be measured with
relatively small aperture radii (5\farcs5, 5\farcs6, and 10\farcs5 for 70, 100,
and 160 \micron, respectively), while brighter objects should be measured with
larger aperture radii (12\arcsec, 12\arcsec, and 22\arcsec\ for 70, 100,
and 160 \micron, respectively).  The larger apertures are usually big enough
to accurately measure the partially resolved objects, even for $z<0.05$, at the
same time still able to avoid contaminating sources and minimize noise.
For highly resolved objects, they chose to use more extended apertures of radii
18\arcsec, 18\arcsec, and 30\arcsec\ for 70, 100, and 160 \micron,
respectively.  The same method is used to determine the aperture of the QSO2s.
Since SDSS images are available for the entire sample, we further examine
the optical sizes of the sources.  We extract and compare their fluxes
densities using the small, large, and, if necessary, extended aperture.  Only
a few objects in our sample need to be measured with larger apertures.  For
example, SDSS J1200$+$3147 and SDSS J1238$+$6703 are ongoing mergers whose gas
distributions are likely to be complex and distributed on multiple scales.
For most of the objects, the variation of flux density as a function of
different aperture sizes is always consistent within $1-2\, \sigma$ of the
sky variation, proving that the aperture sizes are properly determined.  The
listed flux densities account for contamination by companions, as described
below.  To determine the uncertainties of the targets, we perform 20
measurements, centering the apertures evenly on the background annulus (with
radius 45\arcsec) without background subtraction.  The aperture sizes are
exactly the same as those used to measure the sources.  The standard deviation
of the 20 measurements constitutes the $1\,\sigma$ uncertainty of the aperture
photometry of the source \citep{Balog2014ExA,Shangguan2018ApJ}.

A few quasars have bright, close companions, which need to be removed to
minimize their contamination.  We again use \galfit\ to
simultaneously deblend the sources and companions.  In order to generate the
PSF for \galfit, we use the observations of $\alpha$ Tau (obsid: 1342183538
and 1342183541; \citealt{Balog2014ExA}), which we reprocessed with the same
parameters as the quasars.  Visual examination of the residual images show
that the companions are very well removed from the images.  Aperture photometry
for the targets is performed on the residual images with the companions
removed.  Comparing the aperture photometry before and after companion
removal, we find that the extracted flux densities of 10 quasars\footnote{SDSS
J0753$+$3847, J0936$+$5924, and J1356$+$4304 are contaminated in all three
PACS bands. SDSS J1002$+$0551, J1022$+$4734, J1101$+$4004, J1356$+$4259,
J1405$+$4026, J1450$-$0106, and J1605$+$0742 are contaminated in the 160
\micron\ band.} are affected by $>5\%$. We only adopt the measurements from
the contamination-decomposed maps for the 10 objects with the corresponding
bands (marked in Table \ref{tab:herschel}), since the deviation is higher than
the calibration uncertainty (5\%).  SDSS J0753$+$3847, at $z<0.1$, is
well-resolved, and its photometry is measured using the extended aperture.
SDSS J0936$+$5924 resides in a complex region.  It has at least one physical
companion at $z \approx 0.096$, and a group of galaxies at $z \approx 0.04$ is
projected against it. Since the center of the aperture cannot be reliably
determined by fitting the source with a Gaussian profile in the
companion-subtracted maps, we fix the center to the nominal optical
position.  Weak, extended emission may be present at 70 \micron.  We use a
large aperture to enclose all the probable emission, even at the expense of
incurring larger noise.  SDSSJ1356$+$4304, like SDSS J0936$+$5924, also has
a poorly determined centroid after companion deblending; we fix the center of
the aperture but use a small aperture.  The aperture for the rest of the
objects, as for the main sample, is based on their 100 \micron\ flux density.

\subsubsection{SPIRE}
\label{ssec:spire}
The \spire\ imaging photometer (\citealt{Griffin2010AA}) covers a
field-of-view of $4^{\prime} \times 8^{\prime}$ with a FWHM resolution of
$18\farcs1$, $25\farcs2$, and $36\farcs6$ for the 250, 350, and 500 \micron\
bands, respectively.  The observations were conducted in the
small-scan-map mode, with a single repetition scan for each object and
a total on-source integration time of 37 s.

Data reduction was performed using HIPE (version 15.0.1; calibration tree
{\tt spire\_cal\_14\_3}), following pipeline procedures to reduce the small maps.
Although there are several bright targets, many of our sources are faint
($\lesssim$ 30 mJy) and even undetected.  Following the suggested strategy of
photometry for SPIRE, we choose the HIPE build-in source extractor
{\tt Sussextractor} \citep{Savage2007ApJ} to measure the positions and flux
densities of the sources, with the error map generated from the pipeline and
adopting a $3\,\sigma$ threshold for the detection limit.  We measure the source
within the FWHM of the beam around the nominal position of the quasar.  For the
three objects observed by other programs, we measure SDSS~J0843$+$2944 and
SDSS~J1034$+$6001 with the same method. SDSS~J1218$+$0222 is located on the
edge of the map, so we use the measurements from the \herschel/SPIRE Point
Source Catalog (HSPSC; \citealt{Schulz2017arXiv}).  We quote the flux densities
and uncertainties provided by the SUSSEX source extractor.
Following \cite{Leipski2014ApJ}, we use the pixel-to-pixel fluctuations of the
source-subtracted residual map to determine the uncertainty of the flux
measurements.  The residual map is created by subtracting all sources found by
the source extractor from the observed map with $3\,\sigma$ threshold.  We
then calculate the pixel-to-pixel RMS in a box of size 8 times the beam FWHM
of each band.  The box size is large enough to include a sufficient number of
pixels for robust statistics, but small enough to avoid the low-sensitivity
area at the edges of the map.  The median values of the RMS are 10.66, 9.18,
and 11.29 mJy at 250, 350, and 500 \micron, respectively. \cite{Leipski2014ApJ}
found that this method tends to obtain uncertainties very close to, but a bit
smaller than, those calculated from the quadrature sum of the confusion noise
limit and the instrument noise \citep{Nguyen2010AA}.  With a single repetition
scan, the expected noise level is 10.71, 9.79, and 12.76 mJy at 250, 350, and
500 \micron, respectively, very close to our actual measurements.  We provide
$3\,\sigma$ upper limits for all non-detections.  Sources with flux densities
below 3 times the RMS, even if detected by the source extractor, are
considered non-detections.

We visually check the SDSS and PACS images to identify the sources that 
may be contaminated by close companions in the SPIRE maps due to poor 
resolution\footnote{The FWHM of the beam is 18\farcs1, 25\farcs2, and 36\farcs6 
at 250, 350, and 500 \micron, respectively.}.  Except for those undetected in 
the SPIRE bands, the following objects are possibly contaminated: SDSS J0753$+$3847,
J1102$+$6459, J1109$+$4233, J1258$+$5239, J1356$+$4259, J1356$+$1026,
J1358$+$4741, and J1405$+$4026.  The potentially contaminated measurements are
marked in Table \ref{tab:herschel}.  However, since the companion sources are
always much fainter than the quasar in PACS maps, we believe that the
contamination is not significant.  SDSS J1605$+$0742 is falsely detected at
500 \micron\ because it is undetected at shorter wavelengths with higher
resolution and comparable sensitivity.

\section{Results}
\label{sec:results}

\subsection{Stellar Mass}
\label{ssec:sm}

The stellar mass is derived from the $J$-band photometry with a mass-to-light
ratio ($M/L$) constrained by the $B-I$ color, following \citep{BdJ01}

\begin{equation}\label{equ:sm}
\mathrm{log}\,(M_*/M_\odot) = -0.4 (M_J - M_{J, \odot}) - 0.75 + 0.34 (B - I),
\end{equation}

\noindent
where $M_J$ and $M_{J, \odot}=3.65$ \citep{Blanton2007AJ} are the rest-frame
$J$-band absolute magnitudes of the galaxy stellar emission and the Sun,
respectively.  The initial mass function (IMF) is converted from the scaled
\cite{Salpeter1955ApJ} IMF to the \cite{Chabrier2003PASP} IMF by subtracting
0.15 dex \citep{Bell2003ApJS}\footnote{\cite{Bell2003ApJS} provide the
conversion from Salpeter IMF to \cite{Kroupa1993MNRAS} IMF, which is close
enough with the conversion to \cite{Chabrier2003PASP} IMF (e.g.,
\citealt{Madau2014ARAA}).}.  We calculate $M_J$ in two steps.  First, the
original $J$-band flux is subtracted by the emission from the AGN dust torus,
following the results of the SED fitting (Section \ref{ssec:fit}).  Then,
K-correction is applied based on a 5 Gyr simple stellar population model
\citep{BC03} assuming solar metallicity and a \cite{Chabrier2003PASP} IMF.  The
uncertainty of the K-correction, considering the uncertainty of the star
formation history, is $\sim 0.2$ mag.   We adopt a constant color, $B - I = 1.7$
mag, which is typical of the QSO2s in our sample Zhao et al. (2019).  The 
uncertainty of the color-based stellar mass is assumed 0.2 dex 
\citep{Conroy2013ARAA}.  The results are listed in Table \ref{tab:results}.  
As a cross check on our stellar masses, we compare our values with those for 
the 48 objects listed in common in the MPA/JHU 
catalog\footnote{\url{http://wwwmpa.mpa-garching.mpg.de/SDSS/DR7/Data/stellarmass.html}}
based on analysis of SDSS optical spectra; the two sets of measurements are
consistent within a scatter of $\sim 0.18$ dex, close to our assumed uncertainty.
The stellar masses of QSO2s span $M_* \approx 10^{10.3}-10^{11.8}\, M_\odot$.
The stellar masses of QSO1s are estimated from high-resolution optical and
near-IR images \citep{Zhang2016ApJ}, assuming a mass-to-light ratio based on 
\cite{Bell2003ApJS}.  We convert the results based on the Salpeter IMF
to the Chabrier IMF by scaling the stellar mass by a factor of 1.5
\citep{Zhang2016ApJ}.  The uncertainties of the host galaxy stellar masses for
QSO1s are likely larger than 0.2 dex, mainly due to contamination by the bright
nuclei of QSO1s.  Nevertheless, as shown in Section \ref{ssec:cmp}, the median
stellar masses of the two groups are very similar.  This suggests that the
uncertainties of the stellar masses of QSO1s are not likely systematic.

\subsection{SED Fitting}

\subsubsection{Models}
\label{ssec:model}

The IR SED of quasars consists of stellar emission, AGN-heated dust
(e.g., torus) emission, and cold dust emission on galactic scales.  We
fit the SED with a Bayesian Markov chain Monte Carlo (MCMC) method.
We refer to \PaperI\ for the details of the fitting method as well as the 
models for the stellar (BC03) and cold dust (\citealt{DL07}; DL07) emission.  
The FIR emission of the torus drops rapidly beyond $\sim 20\,\micron$ 
\citep{Lyu2017ApJ}.  As the cold dust of the large-scale ISM, which dominates 
the dust mass, emanates mainly at $\gtrsim 70\,\micron$, cold dust masses can
be measured robustly using the DL07 model \citep{Vito2014MNRAS,Shangguan2018ApJ}.
We adopt a new version of the CAT3D model \citep{Honig2017ApJ} to fit the AGN
dust torus emission.  This model considers the different sublimation
temperature of silicate and graphite dust, allowing the model to provide
self-consistently more emission from the hot dust at the inner edge of the
torus.  This inner, hot component had been modeled as an additional blackbody
component in previous models such as CLUMPY \citep{Nenkova2008ApJa,
Nenkova2008ApJb}, as well as in the earlier version of CAT3D \citep{Honig2010AA}.
\cite{Gonzalez2017MNRAS} also recently provide a new set of torus templates
based on the CAT3D model.  In Appendix \ref{apd:torus}, we compare the SED
fitting with the two new sets of CAT3D torus models, along side the median
CLUMPY template obtained from PG quasars \citep{Shangguan2018ApJ}.  We
find that the choice of the torus model has little, if any, effect on the
measurements of cold dust properties \citep{Shangguan2018ApJ,Zhuang2018ApJ}.  We
choose to use the results based on the templates of \cite{Honig2017ApJ}, as they
provide the best overall fits.  The SEDs for the QSO1s have the
advantage of including \spitzer\ mid-IR (MIR) spectra, but, fortunately, the cold dust
masses are not compromised by the use of purely photometry-based SEDs
\citep{Shangguan2018ApJ}.  Although an optional polar wind component is
available for the torus templates \citep{Honig2017ApJ}, we choose the templates
without a wind component because they suffice to fit the MIR {\it WISE}\ data, at
the same time avoiding the need to introduce several additional, poorly
constrained free parameters.  None of the QSO2s in our sample exhibits obvious
signs of jet radiation in the \herschel\ bands from radio-loud sources;
therefore, synchrotron emission is not included in any of the fits.  The
parameters and priors for the three model components are summarized in Table
\ref{tab:pars}.

\subsubsection{Fitting Results}
\label{ssec:fit}

\begin{figure*}[htbp]
\begin{center}
\begin{tabular}{c c}
\includegraphics[height=0.21\textheight]{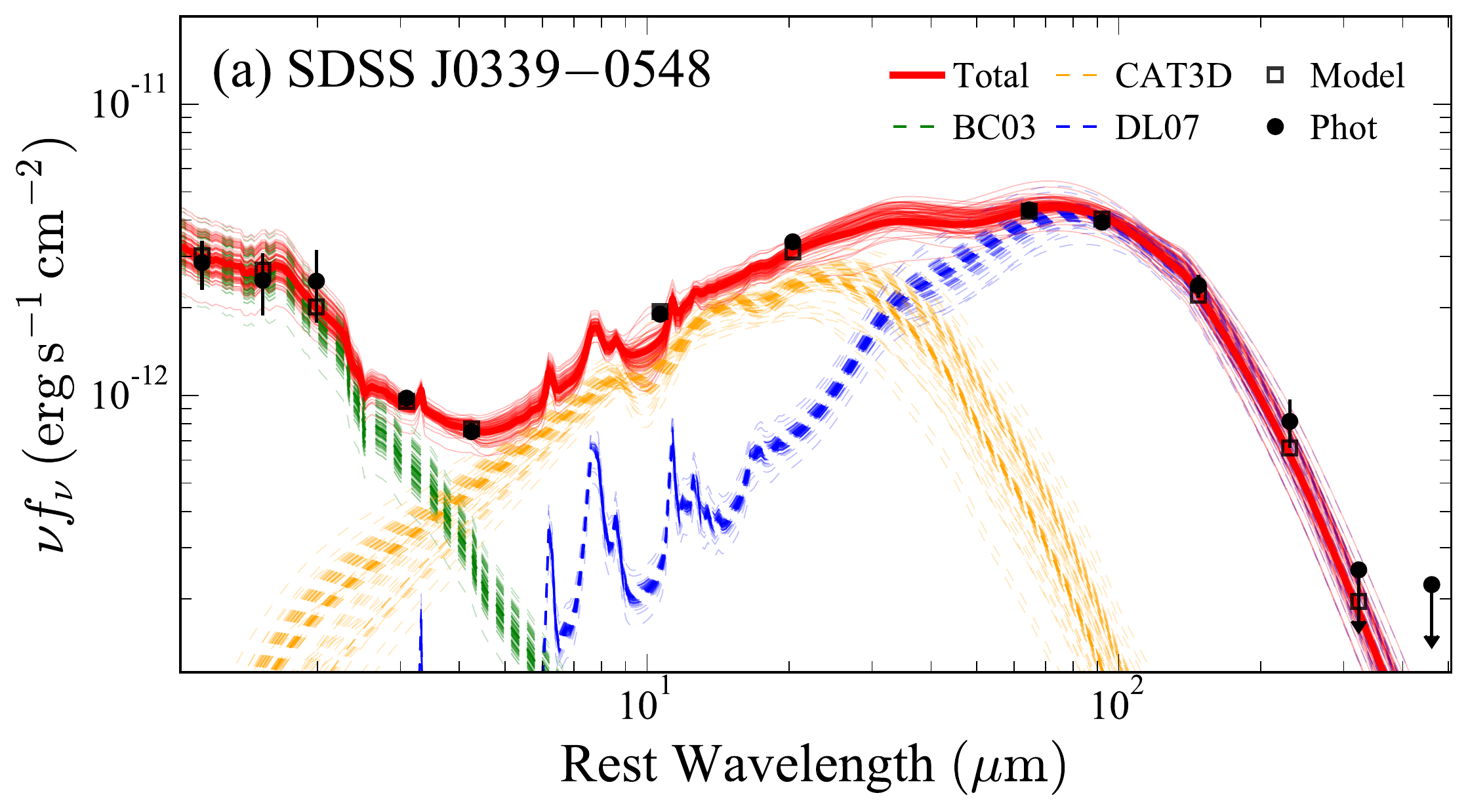} &
\includegraphics[height=0.21\textheight]{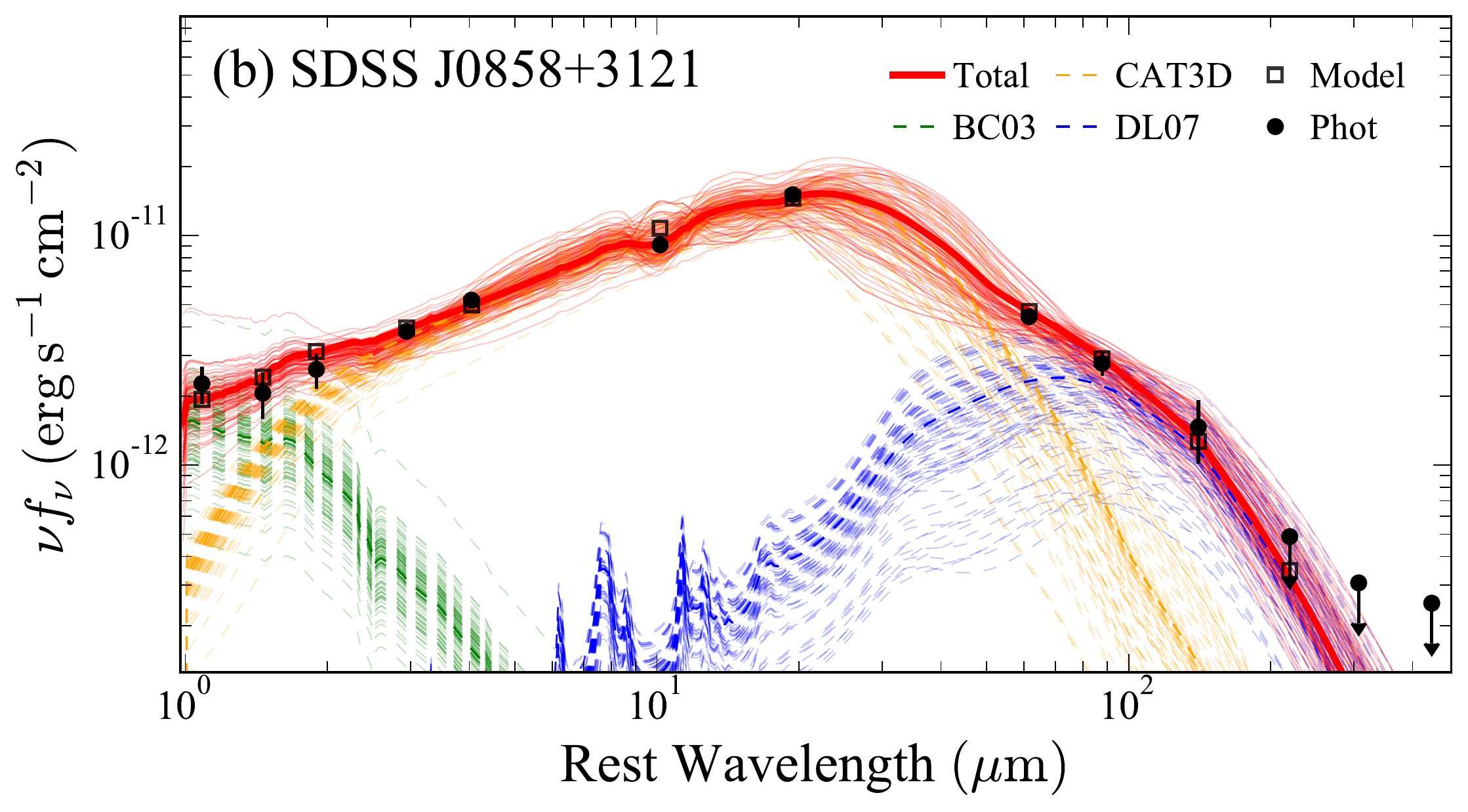}  \\
\includegraphics[height=0.21\textheight]{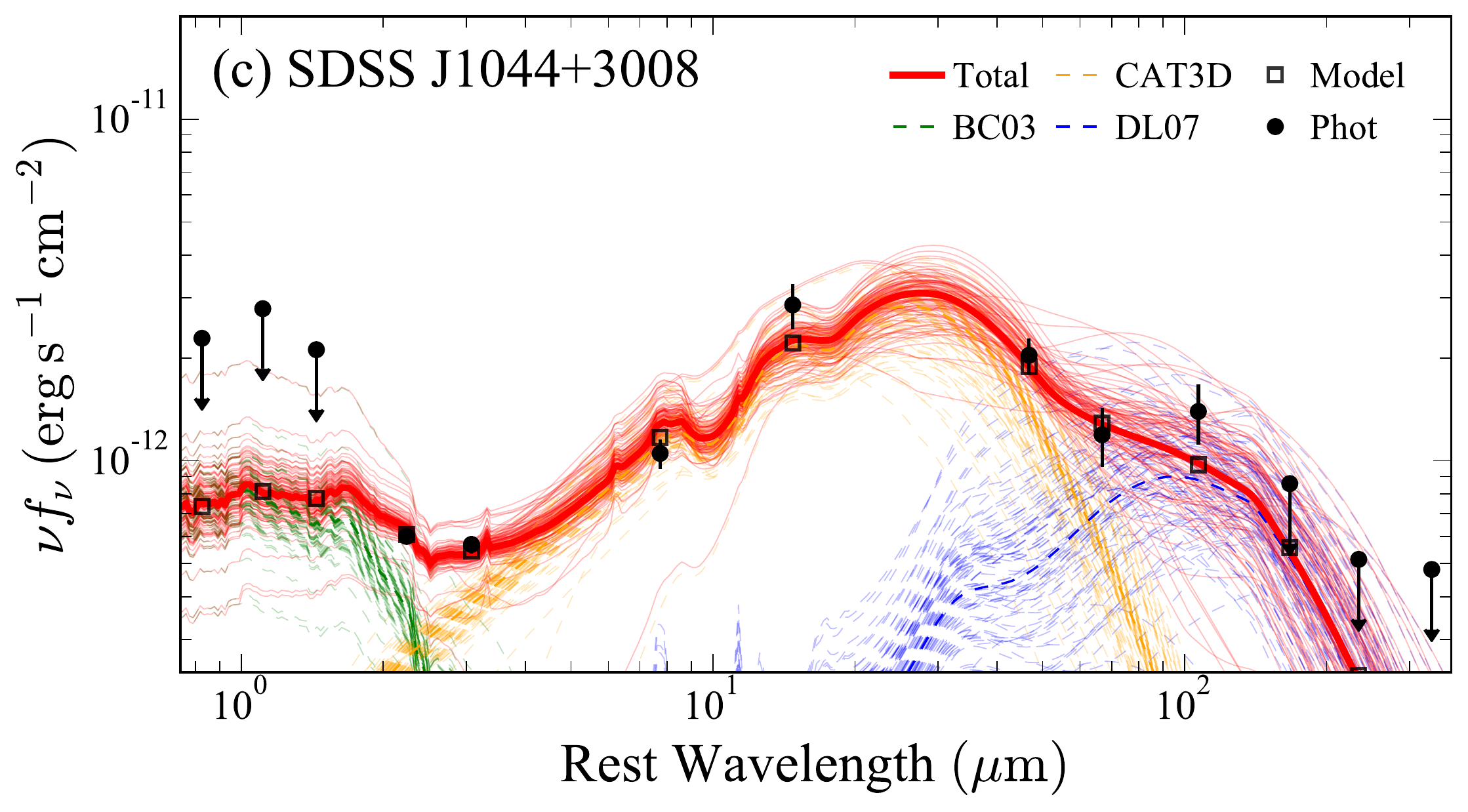} &
\includegraphics[height=0.21\textheight]{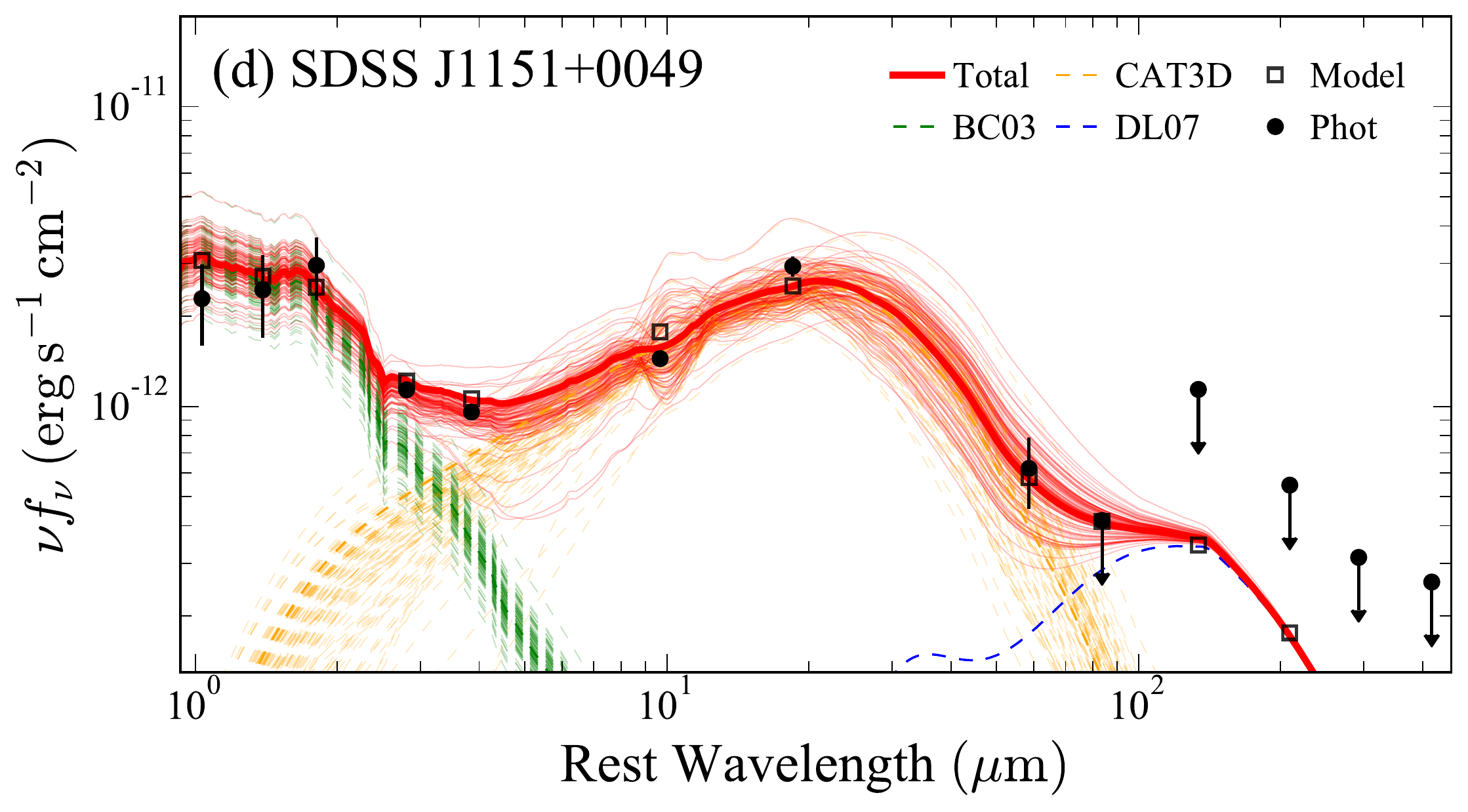}
\end{tabular}
\caption{Four examples of SED fitting results for QSO2s.  The black points are
the photometric data from \tmass, \wise, and \herschel.  The dashed lines are 
the individual components for stars (green; BC03), torus (orange; CAT3D), and 
host galaxy dust (blue; DL07).  The combined best-fit model is plotted as a red 
solid line.  To visualize the model uncertainties, the associated thin lines 
represent 100 sets of models with parameters drawn randomly from the space 
sampled by the MCMC algorithm.  With detections in four \herschel\ bands, 
SDSS J0339$-$0548 (a) has a very well-constrained model.  The fitting for 
SDSS J0858$+$3121 (b) is also reasonably good, because the upper limits provide 
meaningful constraints on the peak of the FIR SED.  SDSS J1044$+$3008 (c) is 
only detected in the \herschel\ \pacs\ bands, and thus the uncertainty of the 
DL07 component is significant.  For SDSS J1151$+$0049 (d), the FIR data are 
even less constraining, and the DL07 component cannot be fit freely.  We fix 
$U_\mathrm{min}=1.0$ and manually adjust the amplitude of the DL07 component 
to estimate an upper limit on $M_d$.  {\it The best-fit results for the entire 
sample of 86 objects can be found in the online version.}}
\label{fig:sed}
\end{center}
\end{figure*}

As shown in Figure \ref{fig:sed}, most of the QSO2 SEDs can be reasonably
well fitted by our combined model.  With more than four bands detected in
\herschel, at least half of the sample can be fitted well because the peak of
the FIR SED is well-constrained by the data (Figure \ref{fig:sed}{\it a}).  Even
with fewer \herschel\ bands detected, it is still possible to constrain the
DL07 model, albeit with relatively large uncertainty (Figures \ref{fig:sed}{\it b}\
and \ref{fig:sed}{\it c}), unless the upper limit at 250 \micron\ is effectively
low.  There are 14 objects detected by \herschel\ with $\lesssim 2$ bands
(e.g., Figure \ref{fig:sed}{\it d}). It is impossible to constrain the DL07 model
for these; instead, the DL07 model, if set free, will mainly fit the mismatch
of the torus component.  Therefore, we fix $U_\mathrm{min}=1.0$ and manually
adjust $M_d$ in steps of 0.1 dex to estimate an upper limit on the dust mass.
Given the same flux, a lower $U_\mathrm{min}$ corresponds to a higher $M_d$.
Note that $U_\mathrm{min}=1$ is the interstellar radiation field in the solar
neighborhood, which is not likely higher than that in a quasar host galaxy.
Therefore, we believe that this is a reasonable assumption to estimate upper
limits for the dust mass.

The torus component of QSO2s is usually much less prominent than that of QSO1s,
mainly because the hot dust emission in the NIR suffers stronger extinction in
obscured AGNs, regardless of where the obscuration comes from.  Interestingly,
five objects\footnote{SDSS J0843+2944, J0858+3121, J1100+0846, J1316+4452, and
J1641+4321.} show strong hot dust emission in the NIR, even prominent enough
to dominate over the stellar component (e.g., Figure \ref{fig:sed}{\it b}).  This
diversity of MIR SEDs is consistent with the analysis of \cite{Hiner2009ApJ}.
The fitting results are listed in Table \ref{tab:results}.

\subsection{ISM Radiation Field}
\label{ssec:isrf}

\begin{figure*}[htbp]
\begin{center}
\includegraphics[height=0.35\textheight]{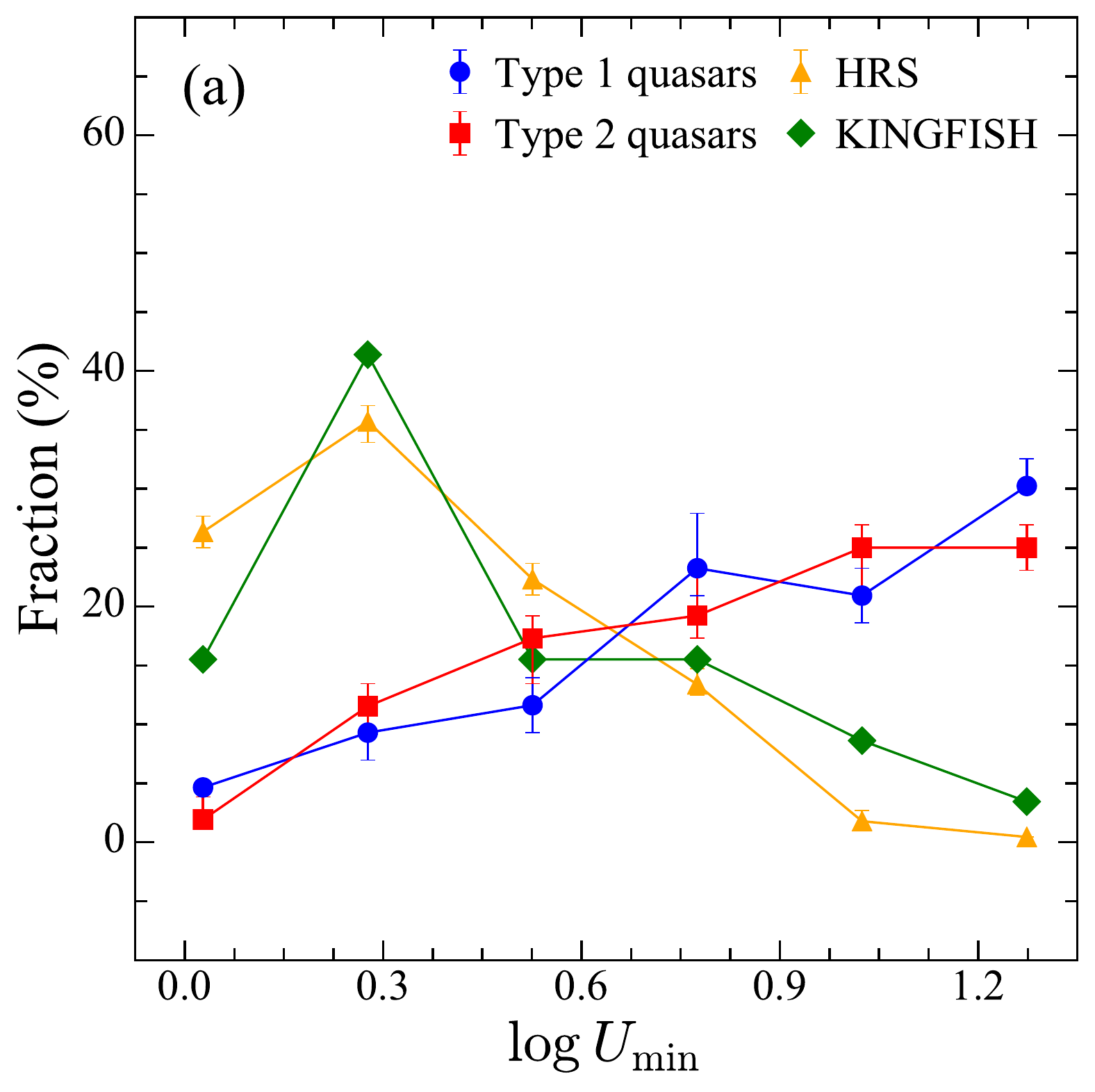}
\includegraphics[height=0.35\textheight]{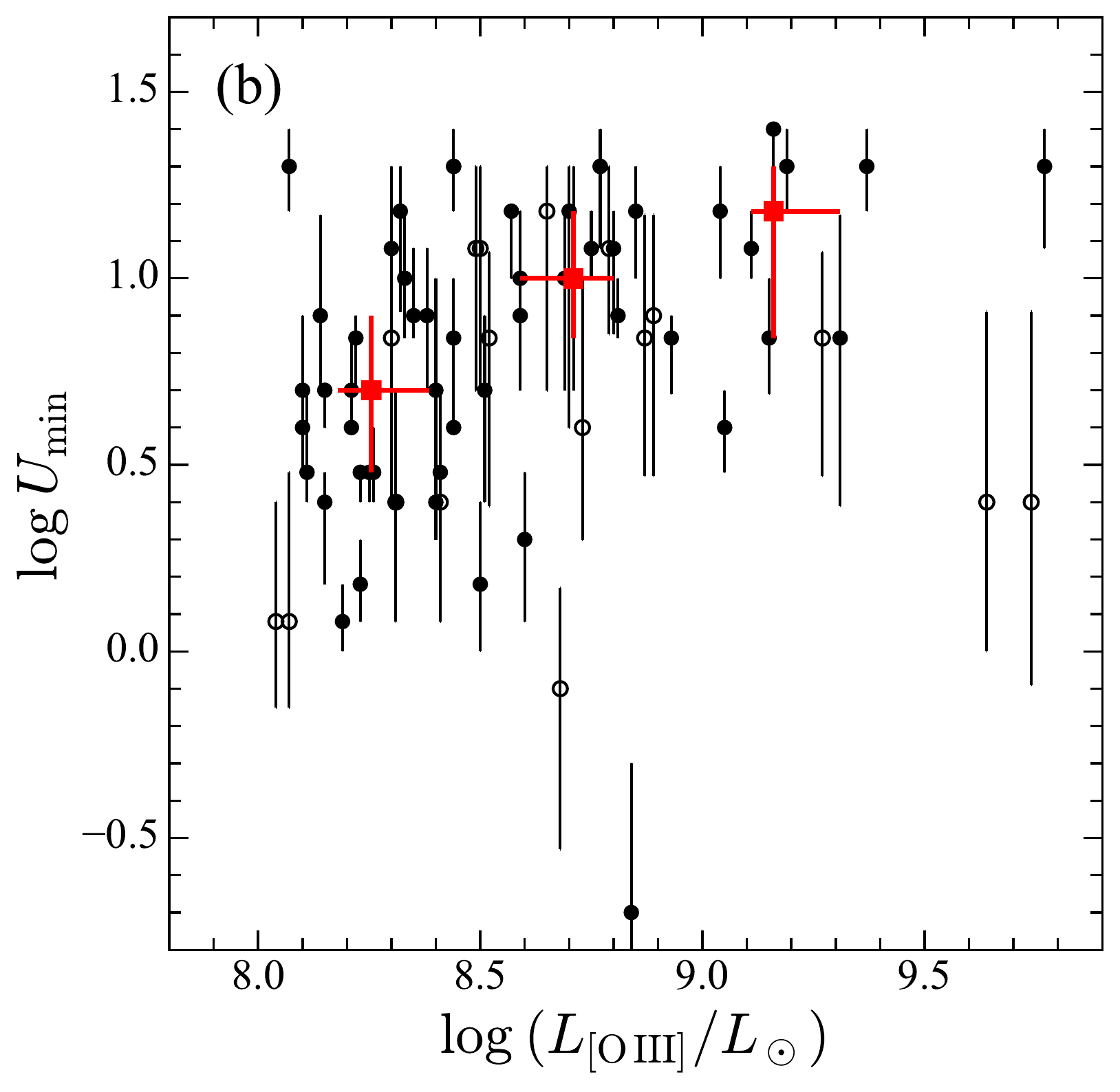}
\caption{The dust-probed interstellar radiation field.  (a) Distribution of
\umin\ for type 1 quasars (blue circles), type 2 quasars (red squares), and
normal galaxies from the KINGFISH (green diamonds) and HRS (orange triangles)
samples.  The errors for the quasars and HRS galaxies are estimated with a Monte
Carlo method, resampling the parameters according to their measured errors and
calculating the number of galaxies in each bin for 500 times.  The error bars
represent 25th--75th percentile ranges of the resampled distribution in each bin.
Since measurement errors of the KINGFISH galaxies are not available, no error
bars are associated with the green diamonds.  The star-forming and quenched
galaxies in the KINGFISH and HRS samples peak at low \umin.  By contrast, the
host galaxies of both types of quasars tend to have higher \umin.  (b) The
values of \umin\ for type 2 quasars generally increase with increasing \OIII\
luminosity.  The filled circles represent more robust measurements than the open
circles; we omitted objects for which only upper limits are available for the
dust mass.  To better visualize the observational trend, we grouped the sample
into three bins: log \loiii\ $< 8.5$, 8.5--9.0, and $\gtrsim 9.0$; the
$50_{-25}^{+25}$th percentile of the distribution in each bin is plotted as red
squares with error bars.
}
\label{fig:dl07}
\end{center}
\end{figure*}

The physical parameter $U_\mathrm{min}$ probes the intensity of the 
interstellar radiation field.   Figure \ref{fig:dl07}{\it a}\ shows the 
distribution of $U_\mathrm{min}$ for QSO2s, comparing them with QSO1s as well 
as normal galaxies from the KINGFISH survey \citep{Draine2007ApJ,
Kennicutt2011PASP} and the Herschel Reference Survey (HRS; 
\citealt{Boselli2010PASP,Ciesla2014AA}).  It is clear that the distributions 
for the host galaxies of {\it both}\ quasar types follow a similar trend, 
rising toward high $U_\mathrm{min}$, while normal galaxies follow the 
opposite tendency, peaking at low values of $U_\mathrm{min}$.  The
large values of $U_\mathrm{min}$ likely indicate that the spatial distribution
of the ISM in both types of quasars is highly concentrated.  Furthermore,
$U_\mathrm{min}$ increases with increasing AGN strength (here taken to be
\loiii, the luminosity of the [O~III] $\lambda$5007 line; Figure
\ref{fig:dl07}{\it b}).  This is consistent with the results from \PaperI, where
$U_\mathrm{min}$ for PG quasars also shows a similar general trend with
increasing AGN continuum luminosity $\lambda L_\lambda$(5100 \AA). The
physical mechanism driving this correlation is unclear.  As discussed in
\PaperI, it might arise from AGN heating of dust in the narrow-line region.
Whatever the exact physical origin, the apparent contribution of AGN heating
to the FIR emission suggests that caution must be exercised in ascribing all
the FIR emission to star formation, as is customary in the literature.

\subsection{ISM Mass}
\label{ssec:imass}

\begin{figure*}[htbp]
\begin{center}
\includegraphics[height=0.35\textheight]{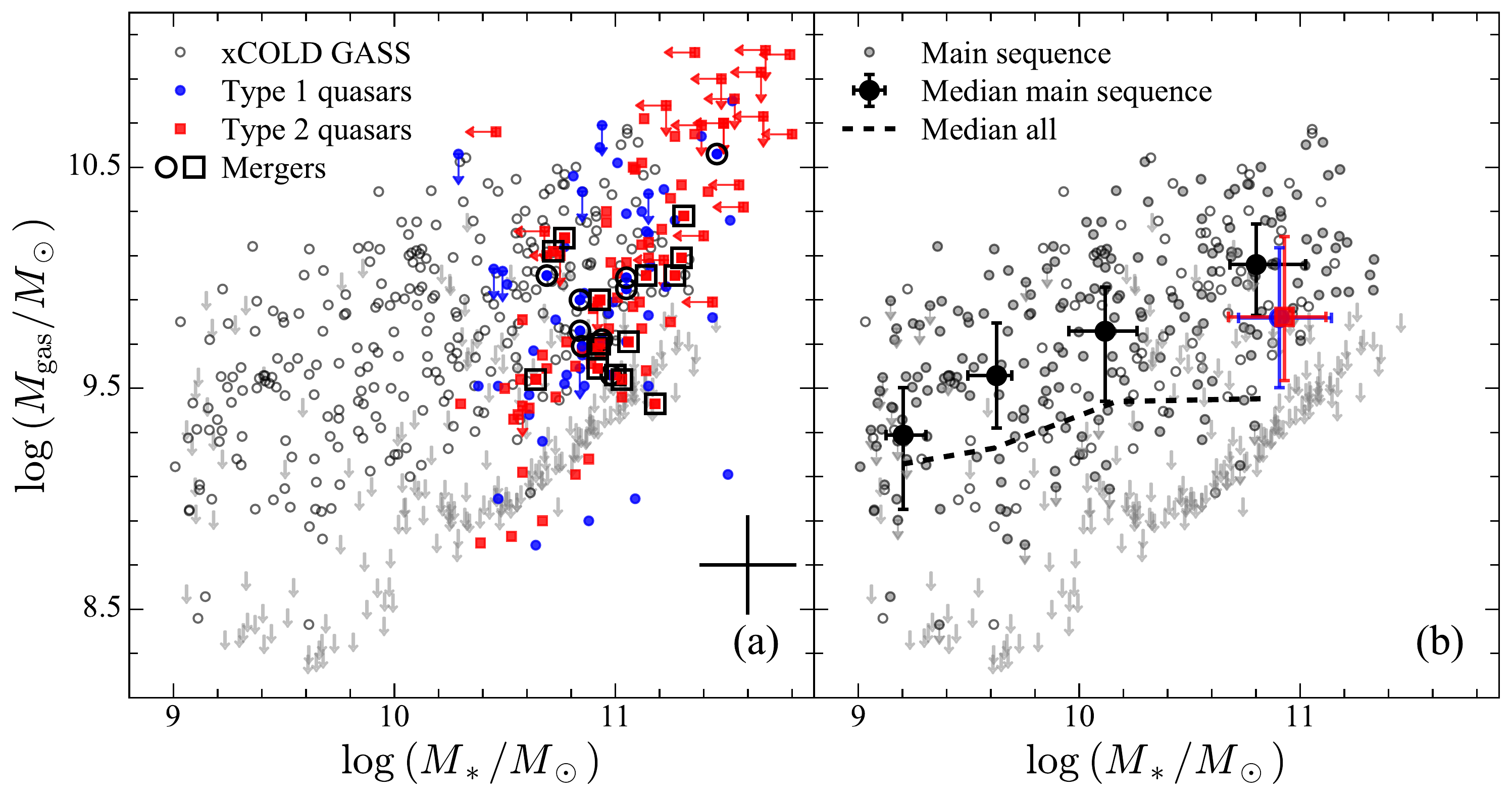}
\caption{The gas and stellar masses of quasars compared to those normal 
galaxies.  The individual objects are plotted in (a), while the medians of the 
quasars and normal galaxies are shown in (b).  Type 2 quasars (red squares) 
display similar distributions of $M_\mathrm{gas}$ and $M_*$ as type 1 quasars 
(blue circles), and both quasar types resemble star-forming galaxies (grey 
circles).  Median gas and stellar masses of the two quasar types are very 
similar.  They are also consistent with those of the normal galaxies on the 
main sequence (individual objects: filled gray circles; median: large black 
circles).  The possible differences between the median gas mass and 
stellar mass of quasars and main-sequence galaxies in the most massive bin are 
well within the error bars.  Meanwhile, quasars show higher gas masses 
compared to the median of all the normal galaxies (dashed line), as many 
gas-poor galaxies are included. Quasars at $z<0.15$ involved in a galaxy merger
[large empty circles and squares in (a)] do not stand out in any particular 
way.  The error bars in (a) illustrate the typical uncertainty of the gas and 
stellar masses for the quasars. The error bars in (b) represent the 25th--75th 
percentile ranges of the sample distribution.}
\label{fig:ism}
\end{center}
\end{figure*}

We convert the dust masses, derived from the SED fitting, to total gas masses
following the method developed by \PaperI:

\begin{eqnarray}
M_\mathrm{gas} = M_\mathrm{H\,I} + M_\mathrm{H_2} = M_{d} \, \delta_\mathrm{GDR, total}, \\
\mathrm{log}\,\delta_\mathrm{GDR, total} = \mathrm{log}\,\delta_\mathrm{GDR} + (0.23 \pm 0.03),
\end{eqnarray}

\noindent
where $\delta_\mathrm{GDR, total}$ is the gas-to-dust ratio (\gdr) estimated from
the galaxy stellar mass and corrected to account for the extended \hi\ gas in the
outskirts of the galaxy.  We estimate \gdr\ from the galaxy stellar mass, 
combining the mass-metallicity relation (\mzr; \citealt{Kewley2008ApJ}) and the
\gdr-metallicity relation (\gzr; \citealt{Magdis2012ApJ}).\footnote{See Equations 
(15) and (16) in \PaperI.}  The values of \gdr\ for the objects with measured stellar 
masses lie in the range $\sim 121-145$, with a median of $122\pm6$, similar to 
those of QSO1s reported in \PaperI.  We adopt the median \gdr\ from the stellar 
mass-detected objects for those whose stellar masses only have upper limits.  
Since the uncertainty on \gdr\ is $\sim 0.2$ dex, this will not introduce significantly 
more uncertainty to the gas masses of the objects without stellar mass measurements 
(usually more distant).  The dust and gas masses of QSO2 host galaxies lie in the 
range $10^{6.7} - 10^{8.9}\,M_\odot$ and $10^{8.8}-10^{11.0}\,M_\odot$, respectively 
(Table \ref{tab:results}).  These photometry-based ISM masses of QSO2s can 
be directly compared to those of QSO1s, as the additional benefit of \spitzer\ spectra 
enjoyed by the latter matters little for the dust masses (Shangguan et al. 2018).

\subsection{Comparison of Gas Masses of Quasars and Galaxies}
\label{ssec:cmp}

As shown in Figure \ref{fig:ism}, the gas content of QSO1s strongly overlaps
with that of QSO2s, and both quasar types possess comparable amounts of ISM as
normal, star-forming galaxies of the same stellar mass.\footnote{The
star-forming galaxies are from the xCOLD GASS survey \citep{Saintonge2017ApJS},
a representative, mass-selected ($M_* > 10^9 M_\odot$) sample of 532 local
($0.01 < z < 0.05$) galaxies with both CO(1$-$0) and \hi\ measurements.}  It is
clear that the median values of the quasars are very close to the median value 
of the normal galaxies within $\pm 0.4$ dex \citep{Chang2015ApJS} around the
star-forming main sequence derived by \cite{Saintonge2016MNRAS}.  Meanwhile, 
quasars are more gas-rich than the entire normal galaxy sample, in which many
gas-poor systems with quenched star formation are included, consistent with
\cite{Vito2014MNRAS}.  The similarity of the cold ISM content of the two quasar
populations is further reinforced by the close resemblance of their median FIR
SEDs (see Section \ref{ssec:msed}).  The quasars hosted by galaxy mergers
that clearly display tidal features (see Section \ref{ssec:sample}) do not show
significant difference in gas and stellar mass distribution compared to the
rest of the sample.  Note that the same conclusions are obtained 
by comparing the dust mass of quasars with that of normal galaxies in the 
HRS sample, although the number of objects with $M_*>10^{10.5}\,M_\odot$ 
is relatively small.  

Sanders et al. (1988) proposed that gas-rich major mergers produce 
ultraluminous IR galaxies, which then evolve into QSO1s after the gas 
and dust are cleared.  While ISM masses exist for local luminous and 
ultraluminous IR galaxies (e.g., \citealt{Shangguan2019ApJ}), comparing 
them to those of quasars is fraught with difficulty because IR-selected 
galaxies are biased toward dusty, and hence gas-rich, objects.  Nevertheless, 
we note that more than 60\% of our QSO2s have total IR (8--1000 \micron) 
luminosities in excess of $10^{11}\,L_\odot$.

We employ survival analysis as implemented in the IRAF.ASURV package
\citep{Feigelson1985ApJ} to include the upper limits of dust, gas, and stellar
masses in the statistical comparisons.  Specifically, we use the Kaplan-Meier
product limit estimator ({\tt kmestimate} task) to calculate the $50_{-25}^{+25}$th
percentile of each physical quantity.  The median and $\pm 25$ percentiles of
the two quasar types are very similar for all the masses (dust, gas, and stars).
In order to consider measurement uncertainties, we use a Monte Carlo method
to resample the dust, gas, and stellar masses 500 times assuming Gaussian
distributions with the mean and dispersion following the measured values and
uncertainties.  We then use the Kaplan-Meier estimator to calculate the medians
of the resampled data.  The two types of quasars remain very similar when
uncertainties are considered (Table \ref{tab:med}).
For the subsamples of quasars hosted by mergers at $z < 0.15$, the median
gas and stellar masses of QSO1s are $10^{9.74}$ and
$10^{10.94}\,M_\odot$, while those of QSO2s are $10^{9.71}$ and
$10^{10.97}\,M_\odot$.  The
differences between QSO1s and QSO2s are small and statistically
indistinguishable for the merger-matched subsamples.  The differences between
the merger subsamples and the whole quasar samples are also small.

Furthermore, we also use the {\tt twosampt} task to test
the null hypothesis that the dust, gas, and stellar masses of the two types of
quasars are drawn from the same parent distribution.  The probabilities of the
null hypothesis are 52.4\%, 59.2\%, and 95.1\%, respectively, for dust, gas,
and stellar masses.  We also use the Monte Carlo method to generate 500
resampled datasets and repeat the two-sample test.  The $50_{-25}^{+25}$th
percentiles of the probability of the null hypothesis are $44.7_{-18.2}^{+24.5}$\%,
$48.3_{-22.6}^{+25.6}$\%, and $57.5_{-21.2}^{+19.9}$\%, respectively.
Therefore, we cannot rule out the null hypothesis that QSO1s and QSO2s
are drawn from the same parent distribution.  
We only report the results from the Peto-Prentice generalized Wilcoxon test,
as suggested by \cite{Feigelson1985ApJ}; the results from the other methods
implemented by the task are entirely consistent.

The Kaplan-Meier estimator is also used to calculate the $50_{-25}^{+25}$th
percentile of the gas and stellar masses of the normal galaxies, as well as
the main-sequence subsample thereof.  For the main-sequence galaxies, only a
small fraction of the gas masses are upper limits, so the results from the
Kaplan-Meier estimator are robust.  However, there are too many upper
limits for the gas masses to estimate the lower end of the distribution for the overall
sample, so that we only plot the median values (dashed line) in Figure
\ref{fig:ism}{\it b}.  We divide the normal galaxies into four stellar mass 
bins: $M_* < 10^{9.4}$,
$10^{9.4} - 10^{9.8}$, $10^{9.8} - 10^{10.5}$, and $> 10^{10.5}\,M_\odot$,
ensuring that there are enough objects in each bin for robust statistics.  The
main-sequence galaxies in the last mass bin ($M_* > 10^{10.5}\,M_\odot$) show
gas and stellar masses consistent with those of the quasars.

\section{Discussion}
\label{sec:discuss}

\subsection{Median SED}
\label{ssec:msed}

\begin{figure*}[htbp]
\begin{center}
\includegraphics[height=0.35\textheight]{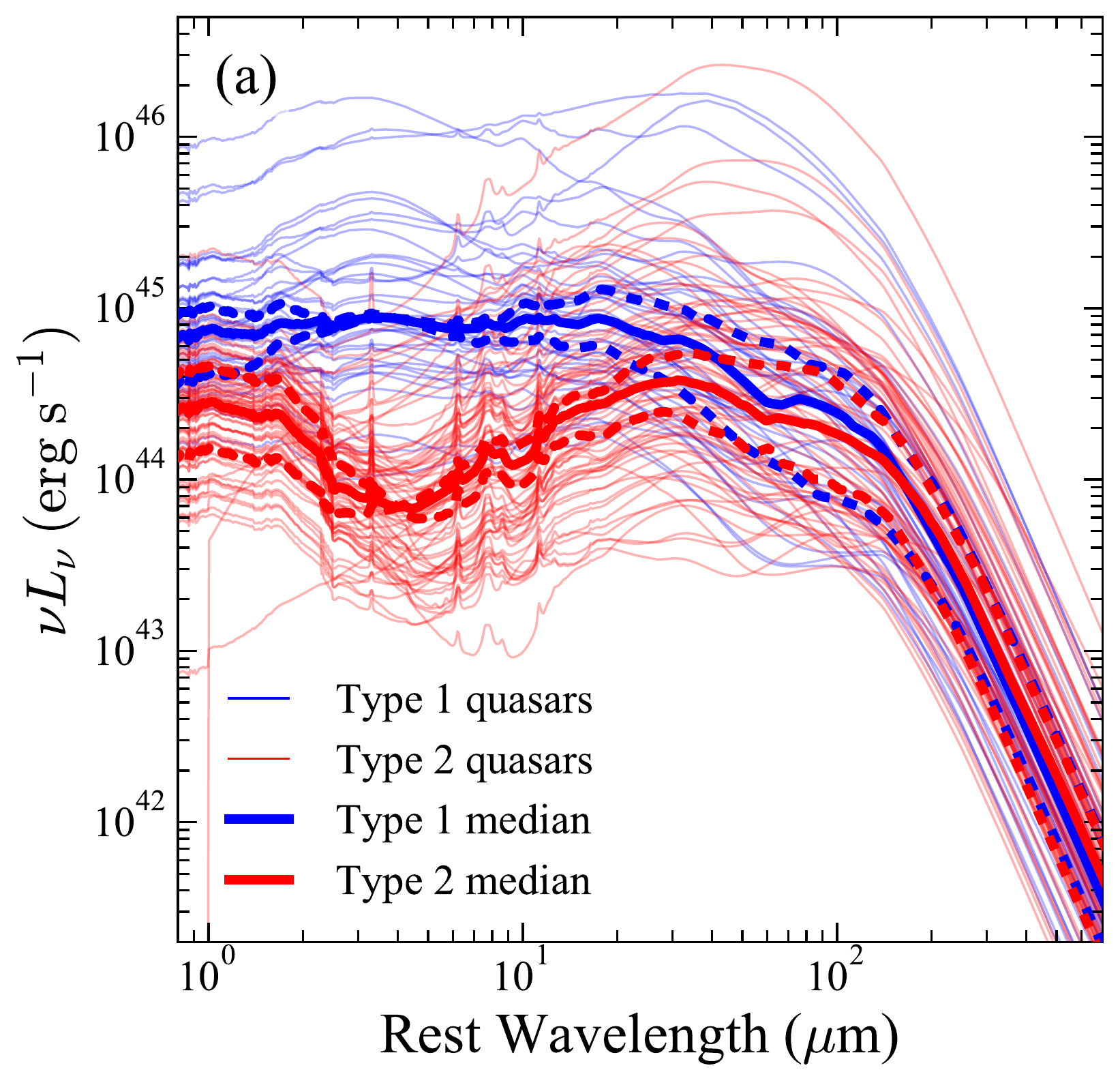}
\includegraphics[height=0.35\textheight]{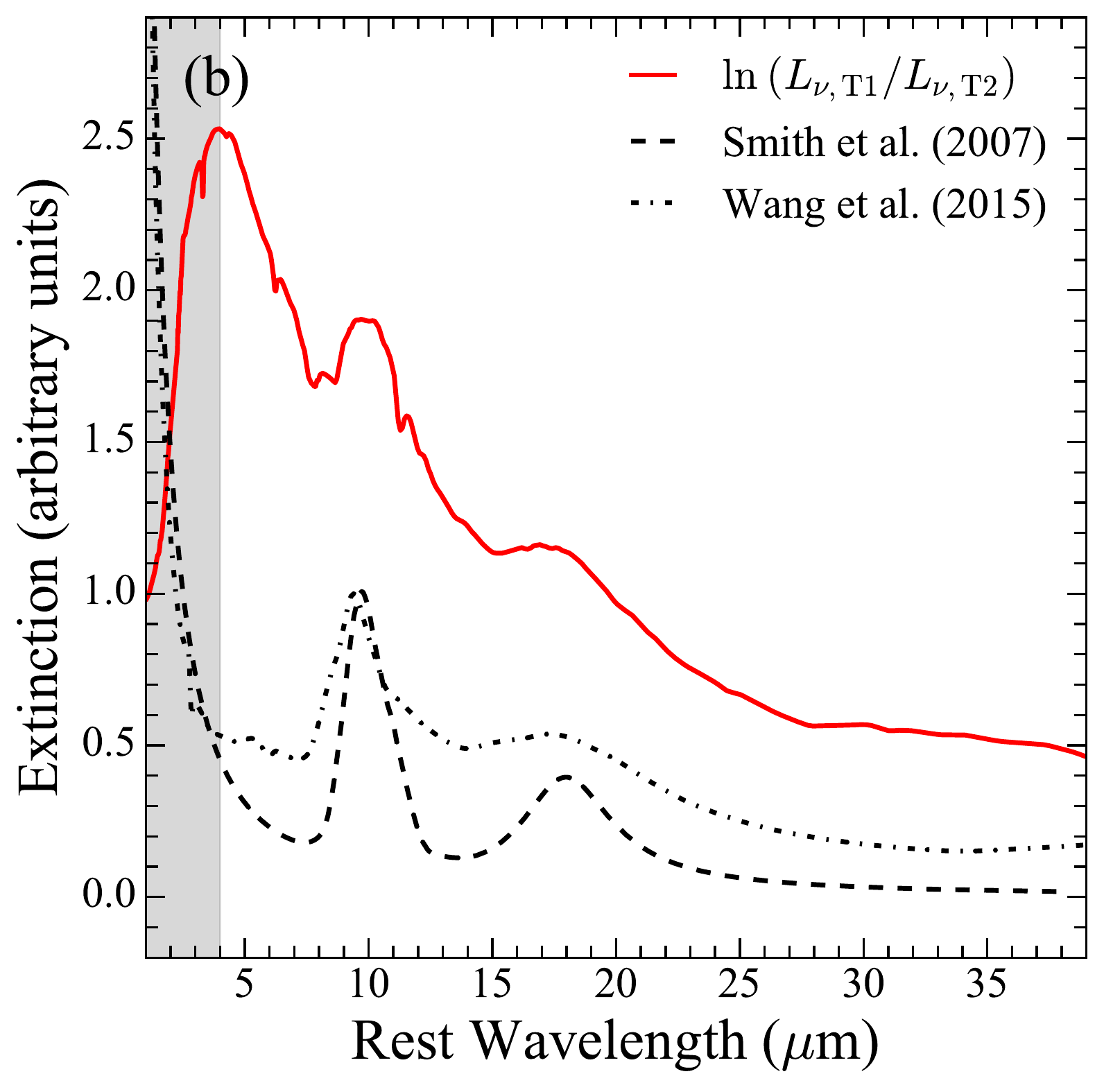}
\caption{Comparison of median SEDs.  (a) The median curves of
type 1 (solid blue curve) and type 2 (solid red curve) quasars are calculated
from the median of the best-fit models (thin lines) normalized at 4 \micron\
and scaled to the median luminosities of the normalization at 4 \micron.
The 25th--75th percentile spread of the distribution is shown as dotted lines.
Type 1 quasars are significantly brighter than type 2 quasars up to $\sim$40
\micron.  At longer wavelengths, the two types have roughly similar SEDs.
(b) The ratio of the median SEDs for type 1 and type 2 quasars (red curve),
which reflects the extinction if the difference between the two median SEDs
is due solely to optically thin dust extinction.  The ratio drops below $\sim$4
\micron\ (shaded region) because stellar emission dominates the near-IR
SED.  The overall rise below 35 \micron\ and the bumps at $\sim$10 and
18 \micron\ are features of the Milky Way extinction curve (e.g., dashed line
and dot-dashed line).}
\label{fig:msed}
\end{center}
\end{figure*}

We generate the median SEDs of QSO1s and QSO2s (Figure \ref{fig:msed}{\it a})
to compare their ensemble properties.  We begin by normalizing the best-fit SED
model of each individual object at 4 \micron, a region with no strong polycyclic
aromatic hydrocarbon emission.  Next, we calculate the median and $\pm 25$th
percentile values of the normalized SEDs at a given wavelength to obtain the
normalized median SED and its dispersion.  In order to compare the absolute
luminosity of the two quasar types, we scale the normalized median SED
according to the median of the luminosity normalizations (at 4 \micron) to get
the final median SEDs.  The quasars for which only an upper limit could be placed
on the dust mass, as well as radio-loud sources, are not included when generating
the median SEDs.  Only the QSO1s with stellar masses are used for the
median SED, but we note that the results are nearly the same if we include the
objects without stellar mass measurements.  The results also hold similarly if
we separate each quasar type into low-redshift and high-redshift subgroups.

Relative to QSO2s, the median SED of QSO1s shows much stronger NIR and MIR
emission.  This is qualitatively consistent with previous works (e.g.,
\citealt{Buchanan2006AJ,Hiner2009ApJ,Hickox2017ApJ}).  The difference in the
shape of the SEDs is as expected from the simple picture that the NIR and MIR
emission from the hot dust surrounding the central engine is less obscured in
QSO1s than in QSO2s.  The FIR components of the two median SEDs overlap closely,
confirming our finding that the cold dust properties of the host galaxies 
of the two quasar types are, on average, very similar.\footnote{We are wary
that the median FIR SEDs may suffer some bias, as we exclude objects with 
\herschel\ non-detections.  However, the bias, if any, likely may not be 
critical, as both quasar samples were observed in exactly the same manner by 
\herschel.} Some earlier works, by contrast, reached different conclusions 
(e.g., \citealt{Hiner2009ApJ,
Chen2015ApJ}), finding that the hosts of QSO1s tend to be less FIR-luminous
than those of QSO2s.  This discrepancy may arise from two reasons.  First, the
MIR selection of the quasar samples in these previous studies may be biased in
favor of intrinsically brighter QSO2s.  Although \cite{Chen2015ApJ} matched
their type~1 and 2 samples based on their best-fit intrinsic AGN bolometric
luminosities, it is still possible that the MIR contribution from the hosts of
QSO2s is higher than that from QSO1 hosts, as the torus emission of QSO2s is
intrinsically more obscured than that of QSO1s.  Second, our quasars lie at
lower redshifts ($z<0.5$) than those of Chen et al. ($0.7<z<1.8$), and thus
redshift evolution may play some role.

Figure \ref{fig:msed}{\it b}\ plots the ratio of the median SED of QSO1s to that of
QSO2s, along side for comparison the Milky Way extinction curves of
\cite{Smith2007ApJ} and \cite{Wang2015MNRAS}.  The SED ratio rises much more
steeply toward short wavelengths than either extinction curve.  The SED ratio
also exhibits discrete bumps at $\sim$10 and 18 \micron\ bumps, which
correspond
to silicate features in the extinction curve.  These features in the SED ratio
may not be robust in detail because the QSO2s in our sample lack spectroscopic
coverage in the MIR.  The clear, monotonic excess toward shorter wavelengths
in the SED ratio indicates that the ISM of quasar hosts is highly optically
thick in the NIR, such that the obscuration does not arise solely from
galactic \citep{Lacy2007ApJ,Hickox2017ApJ} or circumgalactic \citep{Bowen2006ApJ,
Prochaska2014ApJ,Johnson2015MNRAS} scales but instead is more likely highly
concentrated on nuclear scales \citep{Ricci2017Natur}.  While the merger-driven
evolutionary scenario naturally predicts that quasars should have a highly
centrally concentrated distribution of ISM during their obscured (type~2) phase,
we must emphasize that this expectation is at odds with the fact that the
derived distributions of $U_\mathrm{min}$ for QSO1s and QSO2s are remarkably
identical (Figure \ref{fig:dl07}).  The overall spatial distribution of the ISM,
at least as crudely probed by the $U_\mathrm{min}$ parameter of the DL07 dust model,
appears to be quite insensitive to the quasar type.

\subsection{Implications for the Evolutionary Scenario of Quasars}
\label{ssec:imp}

Our central, underlying thesis posits that the major merger-driven, AGN
feedback-mediated evolutionary picture must leave an imprint on the gas content
of the constituent host galaxies.  If ``quasar-mode" AGN feedback is as
effective as suggested by many numerical simulations (e.g.,
\citealt{DiMatteo2005Natur,Costa2018MNRAS}; however, see \citealt{Debuhr2012MNRAS}),
then the cold gas content of the host galaxy, along with its associated dust, is
a key observable parameter that should reflect the evolutionary state of the
system.  In the absence of gas replenishment by significant external accretion
during a merger episode, it stands to reason that QSO1s, as the direct by-product
of dust clearing during the AGN ``blow-out" phase, should have a lower cold gas
content than QSO2s, their immediate, highly obscured precursors.  By the same
token, both quasar types should also be, on average, more gas-deficient than
their progenitor gas-rich galaxies.

\begin{figure}[htbp]
\begin{center}
\includegraphics[height=0.35\textheight]{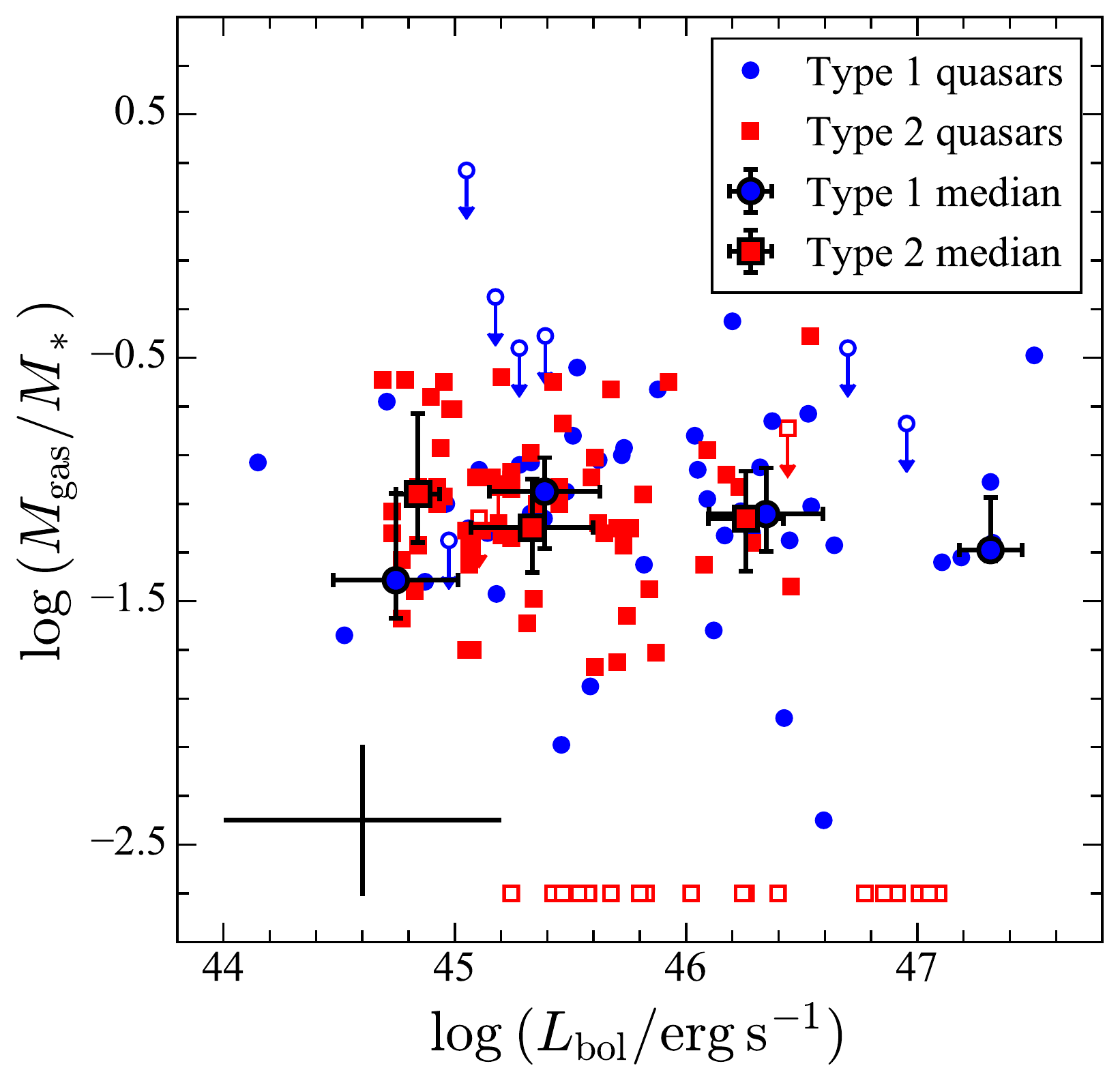}
\caption{The gas fraction does not depend on the bolometric luminosity of 
the quasars.  The quasars with gas mass upper limits are shown with empty 
symbols.  The median gas fractions of the quasars of each type are 
calculated with $L_\mathrm{bol}$ in the ranges of $<10^{45}$, 
$10^{45}$--$10^{46}$, $10^{46}$--$10^{47}$, and $>10^{47}\,L_\odot$, accounting
for the upper limits.  The error bars represent the 25th--75th percentile 
ranges of the data in each bin.  Type 2 quasars with upper limits in both gas 
and stellar masses are assigned an arbitrary mass ratio on the bottom of the 
plot, and they are not included in the calculation of the median gas 
fractions.  The error bars on the lower-left corner illustrate the typical 
uncertainties of the bolometric luminosity and the gas fraction for the two 
types of quasars.}
\label{fig:gdef}
\end{center}
\end{figure}

These simple expectations are not borne out by the observations presented here.  
As discussed in Section 3.4, not only do obscured and unobscured quasars have 
comparable gas content and gas mass fraction, but both types are also essentially
indistinguishable from normal, star-forming galaxies of the same stellar mass.
Figure \ref{fig:gdef} further examines the dependence of the gas mass fraction
on AGN bolometric luminosity.  By design, the two types of quasars in our study
span a similar range of AGN bolometric luminosity.  We estimate the bolometric
luminosity of the quasars using their \OIII\ luminosity as a proxy, following
the formalism of \cite{Stern2012MNRAS}, 
$\mathrm{log}\,(L_\mathrm{bol}/\mathrm{erg\,s^{-1}}) = 1.39\,\mathrm{log}\,
(L_\mathrm{[O\,III]}/\mathrm{erg\,s^{-1}}) - 13.17$.  Although the bolometric
luminosities have substantial uncertainties ($\sim 0.6$ dex), as do the gas 
mass fractions ($\sim 0.3$ dex), it is very surprisingly that these two 
parameters seem to be totally unrelated to each other.  The median gas 
fractions of the two quasar types are consistent with each other within the error 
bars.\footnote{There are 7 (17), 22 (42), 18 (8), and 5 (0) QSO1s (QSO2s) 
used to calculate the median gas fractions with $L_\mathrm{bol}$ in $<10^{45}$,  
$10^{45}$--$10^{46}$, $10^{46}$--$10^{47}$, and $>10^{47}\,L_\odot$ bins, 
respectively.}  This was already noted by \PaperI\ in the case of QSO1s. Here 
we reinforce the same conclusion for a matched sample of QSO2s.

\cite{Ho2008ApJ,Ho2008ApJS} addressed these very issues using \hi\
observations of type~1 AGNs.  They, too, failed
to find any evidence for a deficiency of gas among AGNs compared to a control
sample of inactive galaxies matched in luminosity and morphological type.
However, the majority of the AGNs studied by Ho et al. were quite nearby ($z
\lesssim 0.1$) because of current limitations with \hi\ observations and have
luminosities too low to be deemed bona fide quasars.  By contrast, the current
study and that of \PaperI\ target AGNs sufficiently powerful
to constrain the canonical merger/feedback paradigm.

How to interpret these results?  Taken at face value, they seem to seriously
challenge the popular major merger-driven evolutionary scenario for
transforming gas-rich, inactive galaxies into active systems on the one hand,
and obscured (type~2) AGNs into unobscured (type~1) AGNs on the other.  Perhaps
galaxy mergers trigger quasar activity, but
there is little direct evidence that quasar-mode feedback links the two types
of quasars, or that the feedback controls the overall evolutionary cycle from
normal to active and eventually to quenched galaxies.  
AGN-driven outflows likely recycle gas inside the galaxy instead of eject it out (e.g.,
\citealt{Bischetti2019,Fluetsch2019MNRAS}).  Contrary to our results,
\cite{Perna2018AA} find that the molecular gas fraction of high-redshift
($z>1$) obscured AGNs tends to be lower than that of main-sequence galaxies with
similar mass at corresponding redshift.  Quasar-mode feedback may be more 
effective in the high-redshift Universe when black hole accretion was more intensive.
 
It is unlikely that feedback only pushes the dusty ISM out of the galactic
nucleus while leaving it still bound to the galaxy, as it would still be
difficult to conspire to keep the interstellar radiation fields of the two 
quasar types so similar (Section \ref{ssec:isrf}).  Nor can the problem be 
evaded by appealing to timing.  On the one hand, although the AGN is possibly 
flickers on a time scale of $\sim 10^5$ Gyr \citep{Schawinski2015MNRAS}, the 
entire episode for the black hole to intensively accrete gas (e.g., Eddington 
ratio $\gtrsim 0.1$) still lasts $\gtrsim 100$ Myr \citep{Marconi2004MNRAS}.  
Fortuitously capturing the quasar just after the ``blow-out" phase seems 
improbable.  On the other hand, since quasar host
galaxies have similar gas fractions as normal, star-forming galaxies, the time
lag between star formation and AGN activity, if both are triggered by the same
merger, should not be long.  Given a gas mass of $\sim 10^{10}\,M_\odot$ at
$M_* \approx 10^{11}\,M_\odot$ (Figure \ref{fig:ism}$b$), a moderate star
formation rate of $\sim 10\,M_\odot\, \mathrm{yr^{-1}}$ \citep{Kim2006ApJ,
Zakamska2016MNRAS} would significantly deplete the gas content of the galaxy in
merely $\sim 1\,\mathrm{Gyr}$.

Finally, the indistinguishable gas content would be understood trivially if the
dichotomy between type~1 and type~2 quasars is merely due to viewing 
angle, instead of evolution.  We must emphasize, however, that this explanation 
is not favored by various statistical differences between the two quasar types 
discussed in Section 1, including star formation rates (e.g., 
\citealt{Kim2006ApJ}), ionized gas velocity fields \citep{Greene2011ApJ}, 
radio continuum properties \citep{Lal2010AJ}, and local environment 
\citep{Villarroel2014NatPh}.  Moreover, our results hold even when the 
comparison between the two quasar types is restricted solely to the subset of
objects triggered by galaxy mergers, whose obscuration cannot be explained 
purely by the traditional AGN unified model \citep{Ricci2017MNRAS}.

\section{Summary}
\label{sec:sum}

In order to test the evolutionary link between QSO1s and QSO2s, we analyze
the complete IR ($\sim 1-500$ \micron) SEDs of a sample of 86 QSO2s matched in
redshift and \OIII\ luminosity with the sample of $z<0.5$ Palomar-Green QSO1s
recently studied in the same manner by \PaperI.  We construct the SEDs
using integrated photometric measurements based on \tmass, \wise, and our own
targeted observations acquired with the PACS and SPIRE instruments on
\herschel.  We use a newly developed fitting method with Bayesian MCMC
inference to model the quasar SEDs, decomposing them
into their constituent components from starlight, AGN-heated warm and hot dust
from a torus, and the cold dust emission from the large-scale ISM of the host
galaxy.  We derive dust masses and constraints on the
intensity of the interstellar radiation field.  The key parameters for the
cold dust component are robust with respect to the choice of models adopted
for the AGN dust torus.  We derive total gas masses from the dust masses, and we
estimate stellar masses from NIR photometry and optical colors.

Our main conclusions are as follows:

\begin{itemize}
\item The host galaxies of QSO2s, with stellar masses $M_* \approx 10^{10.3}-
10^{11.8}\,M_\odot$ (median $10^{10.9\pm0.2}\,M_\odot$), are as gas-rich as
normal, star-forming galaxies of comparable mass, considering the 
25th--75th percentiles of the sample distributions.  They have total dust masses
$M_d \approx 10^{6.7}-10^{8.9}\,M_\odot$ (median $10^{7.7\pm0.3}\, M_\odot$),
which correspond to total (atomic and molecular) gas masses of $M_{\rm gas}
\approx 10^{8.8}-10^{11.0}\,M_\odot$ (median $10^{9.8\pm0.3}\,M_\odot$).

\item The host galaxies of QSO2s have very similar dust/gas content and
dust/gas mass fraction as the host galaxies of QSO1s.  In turn, the ISM
content of both types of quasars are essentially indistinguishable from that
of normal, star-forming galaxies of the same stellar mass.  The global
gas content of quasar hosts also shows no correlation with the bolometric
luminosity of the AGN.

\item The interstellar radiation field of the hosts of both types of quasars is
bear a close resemblance to each other, suggesting a similar spatial 
distribution of ISM.

\item The above results are at odds with the major merger-driven evolutionary
model for the transformation of QSO2s to QSO1s.  Moreover, the overall
similarity between the gas content of inactive galaxies and quasars poses
a serious challenge to the efficacy of quasar-mode ejective feedback in
galaxy evolution.

\item The interstellar radiation field of quasar host galaxies is stronger
than that of normal, star-forming galaxies, and it increases in intensity with
increasing AGN luminosity, suggesting that the AGN heats the large-scale ISM
of the host.

\item The median SEDs of QSO1s and QSO2s are virtually identical in the FIR,
as a consequence of their similar dust content, but differ in the MIR due to
greater extinction among the obscured QSO2s.

\end{itemize}

\acknowledgments

We are grateful to an anonymous referee for helpful comments and 
suggestions.  This work was supported by the National Science Foundation of 
China (11721303) and National Key R\&D Program of China (2016YFA0400702).  JS 
thanks the \herschel\ helpdesk for assistance on \herschel\ data reduction.  He
is grateful to Minjin Kim for useful discussions, Mingyang Zhuang for help with
the dust torus models, Dongyao Zhao and Yulin Zhao for help with the stellar 
mass calculation and host galaxy morphology, and Yanxia Xie and Shu Wang for 
help with the extinction curves.


\appendix

\section{Comparisons with Independent Photometric Measurements}

\subsection{2MASS and WISE}
\label{apd:nir}

\begin{figure*}
\begin{center}
\includegraphics[height=0.25\textheight]{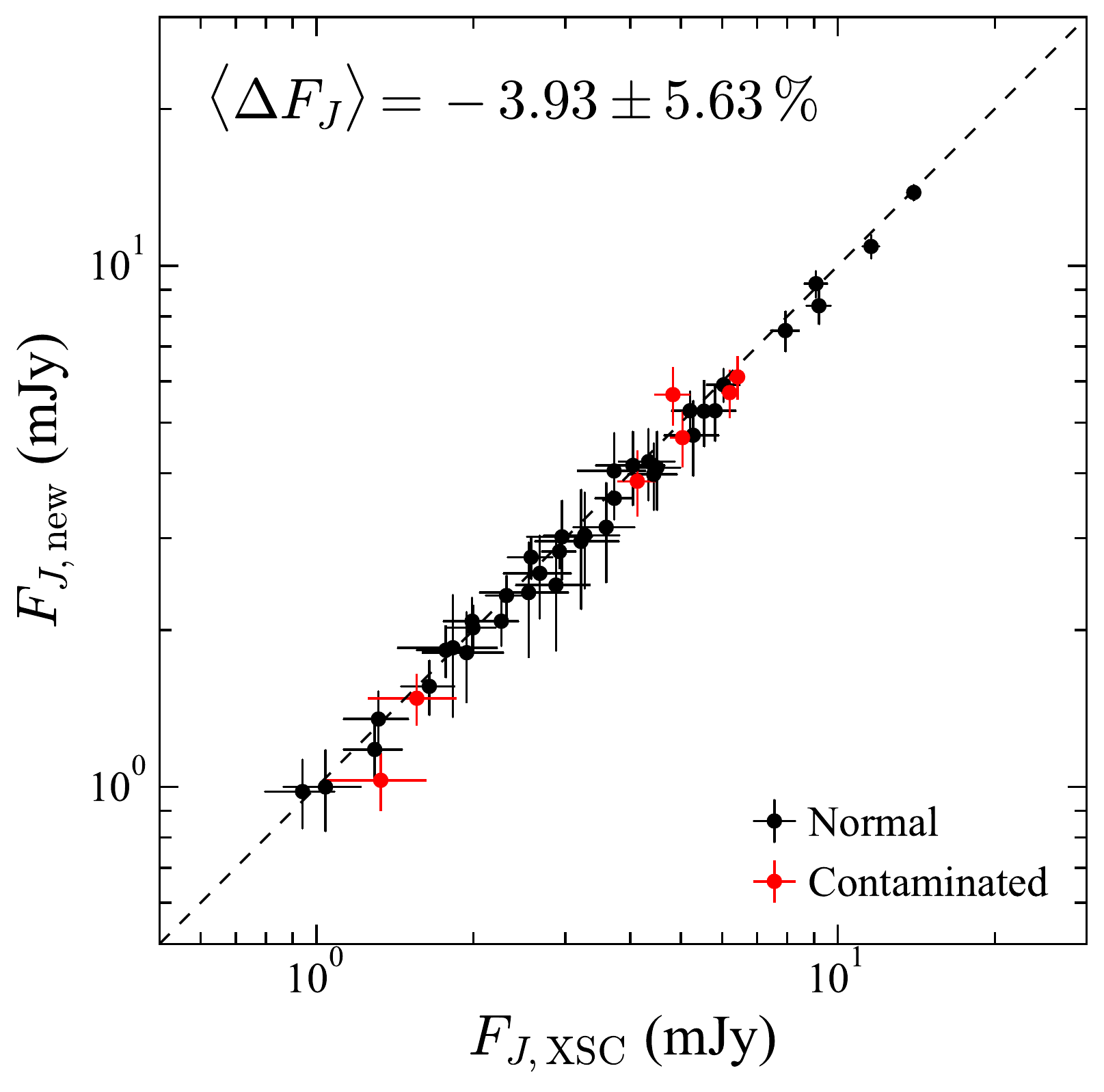}
\includegraphics[height=0.25\textheight]{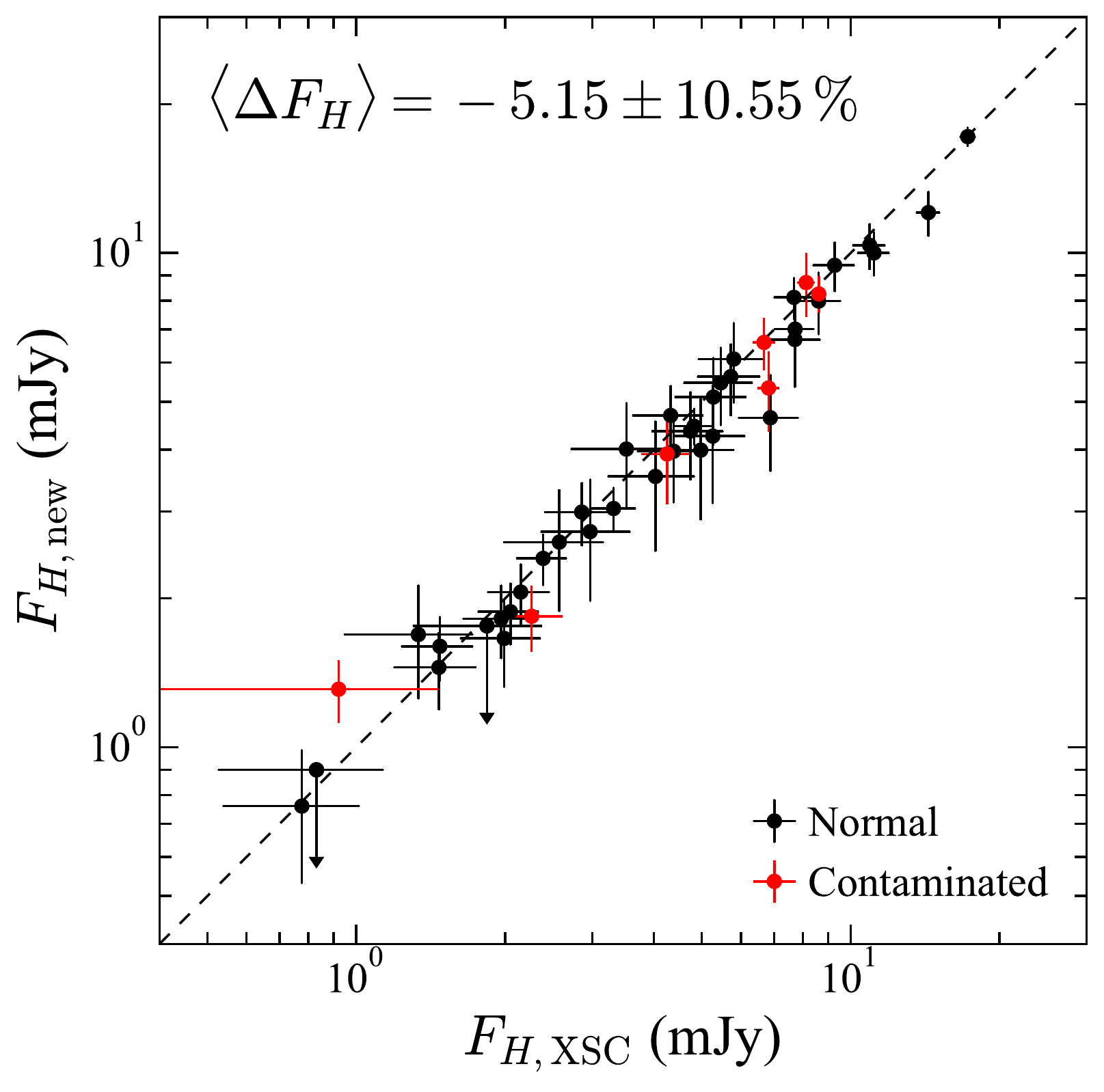}
\includegraphics[height=0.25\textheight]{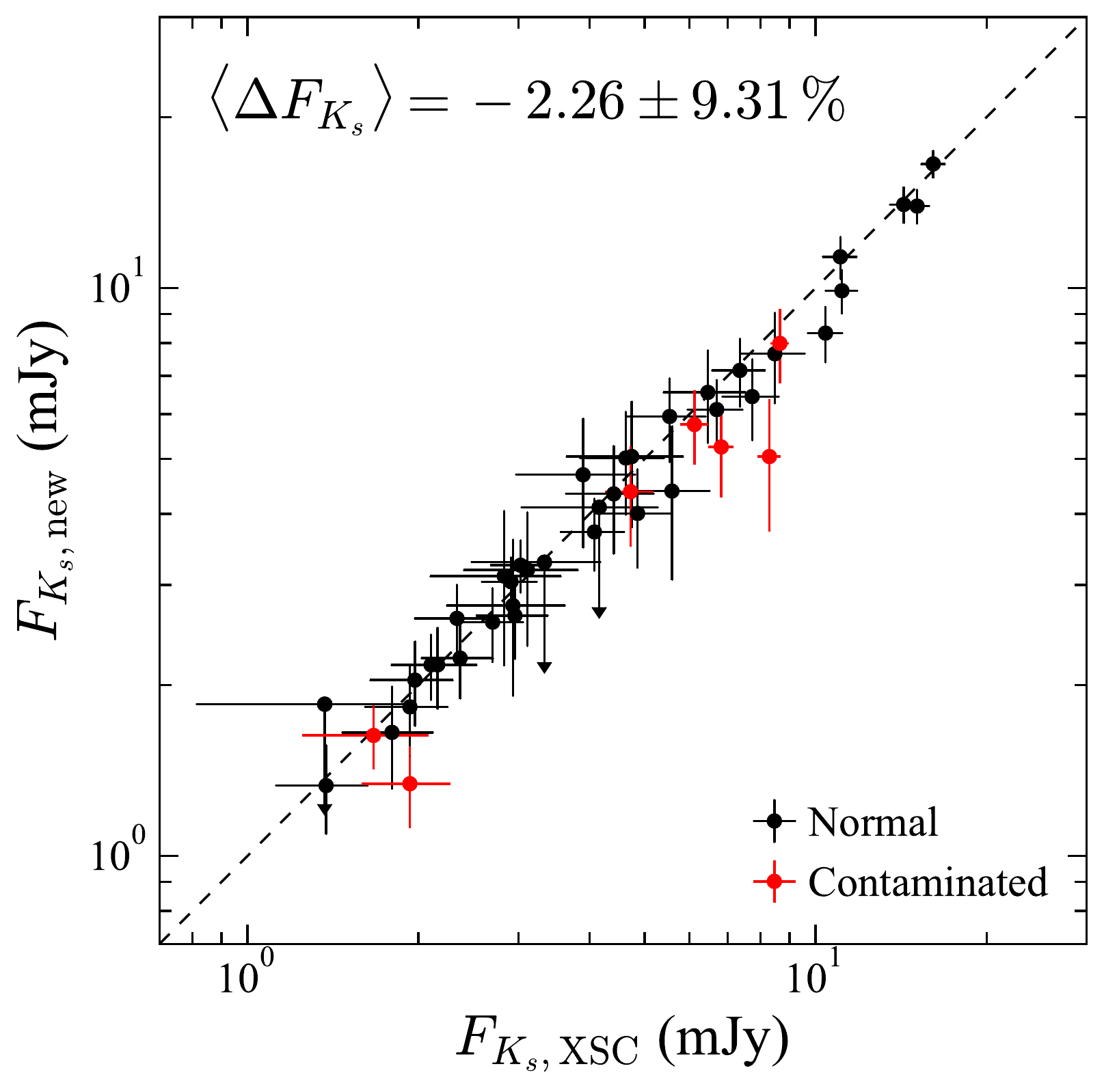}
\caption{Comparison between our new NIR photometry with that from the \tmass\
XCS, for normal (black) and contaminated (red) sources whose companions were
decomposed using \galfit.  $\langle \Delta F \rangle$ is the fractional
deviation in each band.  Our measurements agree well with those in XCS, which
tend to be slightly higher.  Objects with companions or upper limits are not
included in the statistics.}
\end{center}
\label{fig:cmp2m}
\end{figure*}

We compare our aperture photometry with independent measurements listed in the
\tmass\ extended source catalog (XCS; \citealt{Jarrett2000AJ}).  There are
45 objects in our QSO2 sample that match with the XCS using a search radius
of 5\arcsec.  Comparing photometry made with the same aperture radii, our
measurements are reasonably consistent with those in the XSC, with the latter
tending to be slightly ($\lesssim 5\%$) higher than ours (Figure
\ref{fig:cmp2m}). Our results are systematically larger if we were to compare
with the default values in the catalog based on an elliptical aperture set to
a $K_s$-band isophote of 20 $\mathrm{mag\,arcsec^{-2}}$.  This is expected
because the default aperture size is small.

\begin{figure*}
\begin{center}
\begin{tabular}{c c}
\includegraphics[height=0.35\textheight]{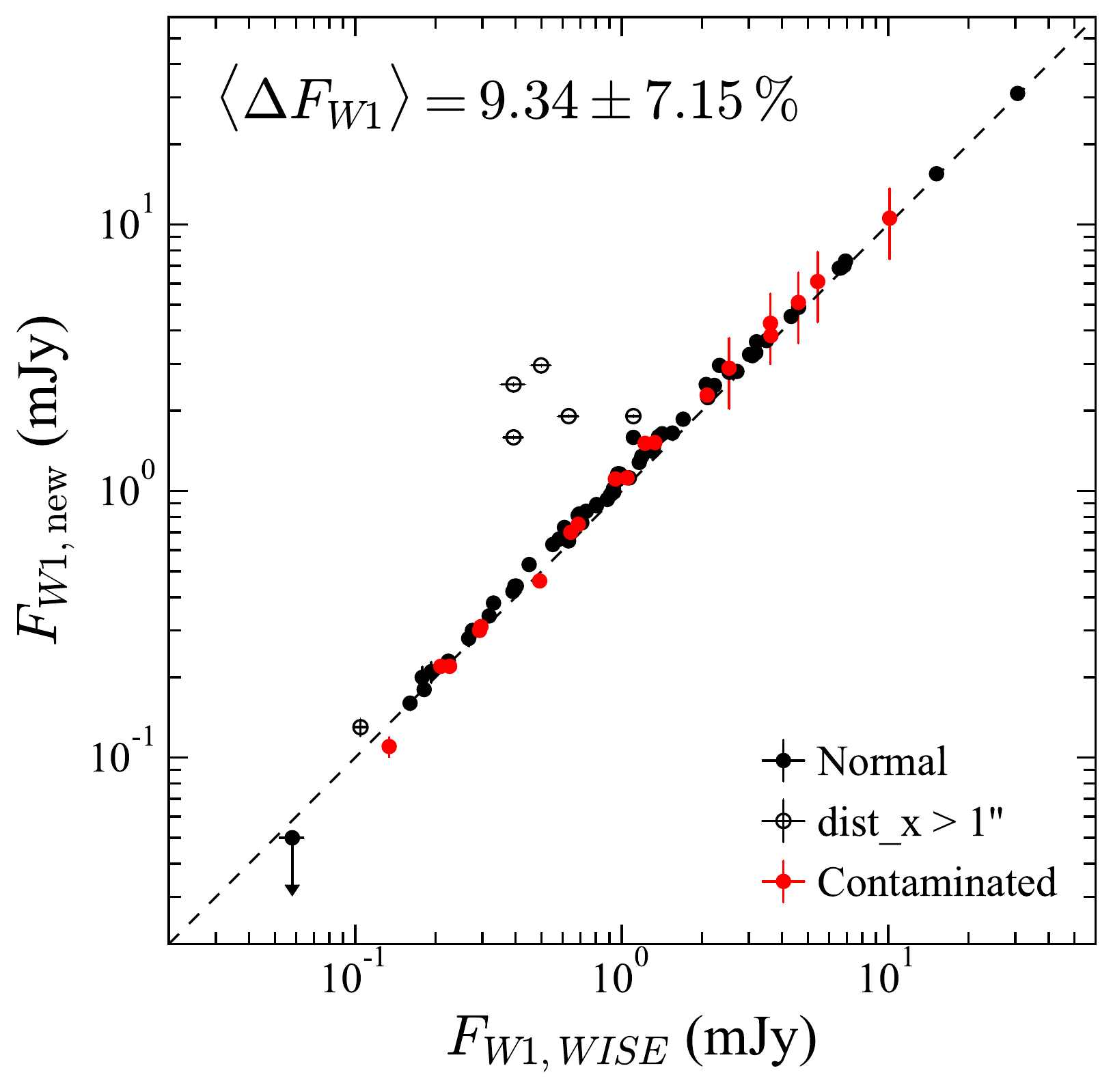} &
\includegraphics[height=0.35\textheight]{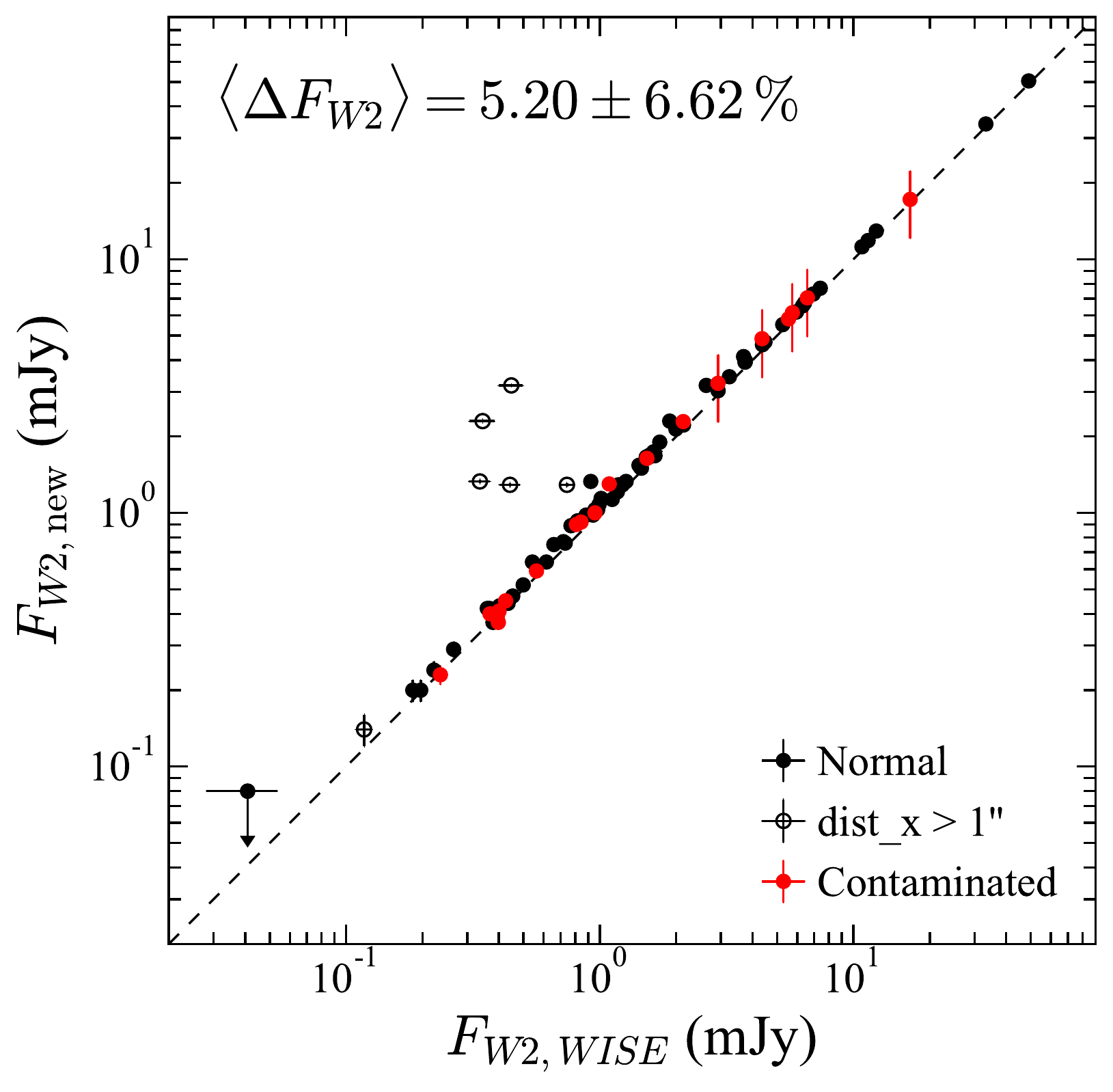} \\
\includegraphics[height=0.35\textheight]{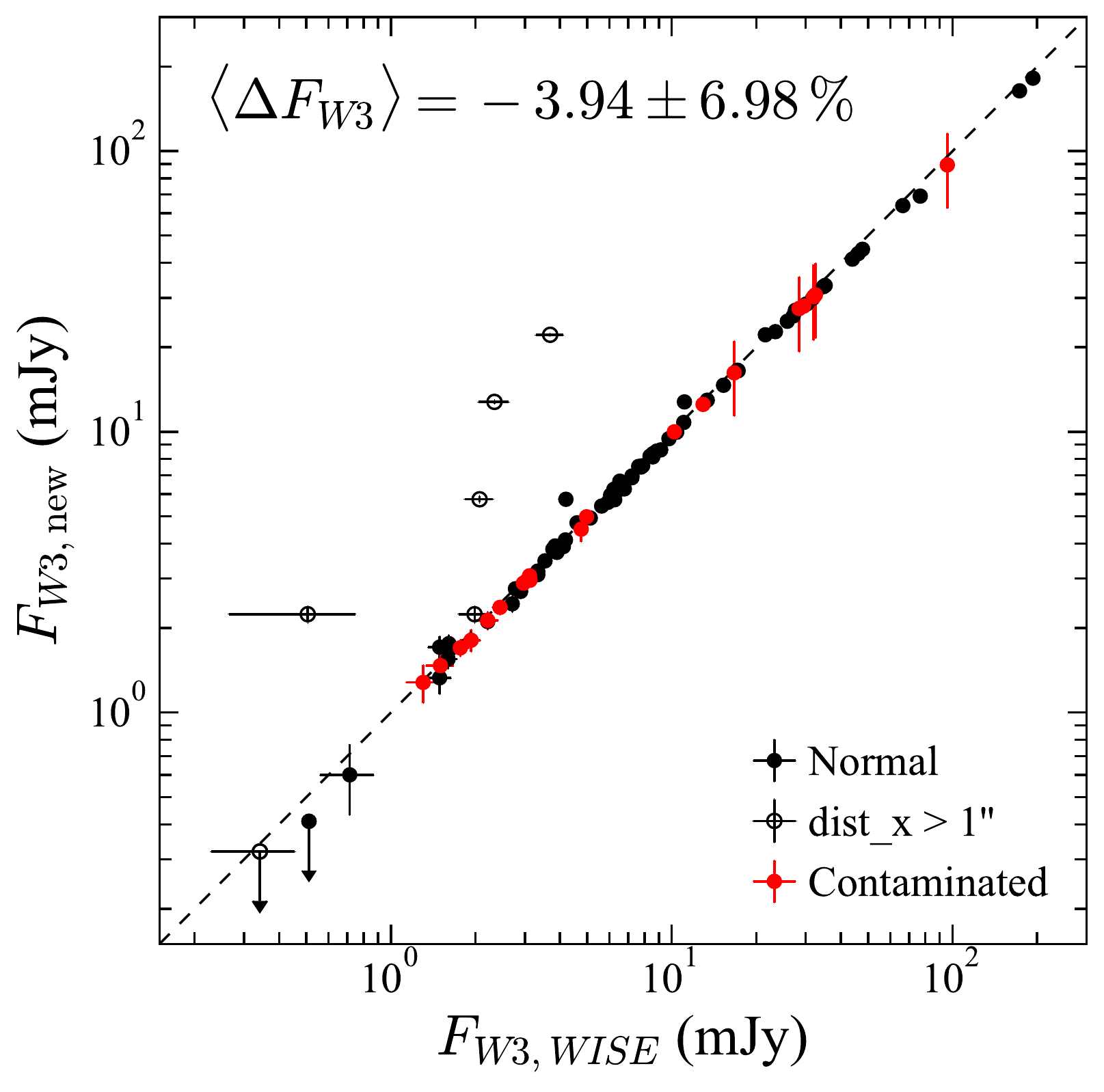} &
\includegraphics[height=0.35\textheight]{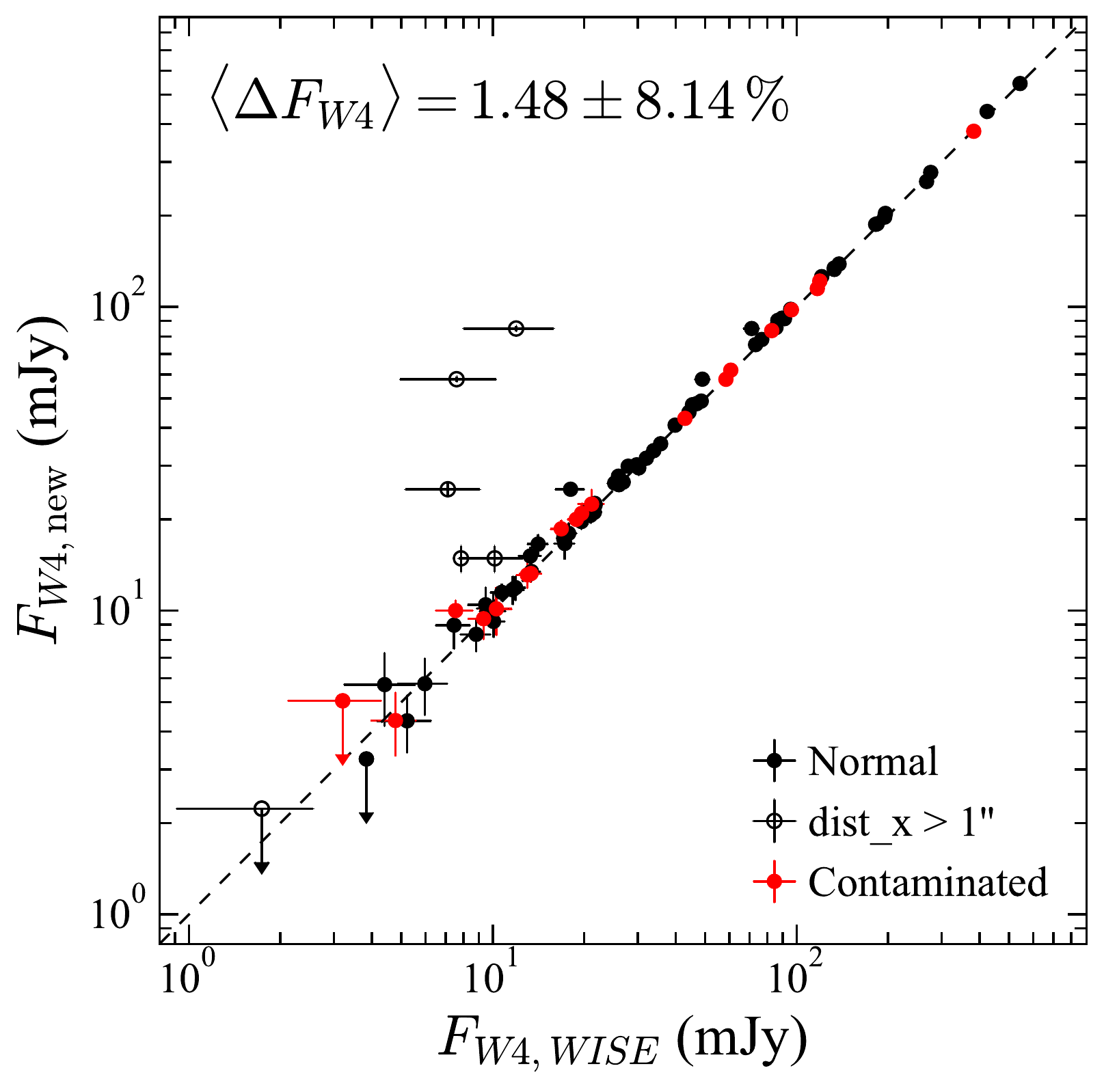}
\end{tabular}
\caption{Comparison between our new MIR photometry with that from the AllWISE
catalog, for normal sources (filled black symbols), contaminated sources with
companions removed using \galfit\ (red symbols), and objects with positions in the
catalog deviating $>1\arcsec$ from our nominal positions (open black symbols).
All sources with atypically large flux deviations are open symbols; they are
likely mismatched.  $\langle \Delta F \rangle$ is the fractional deviation in
each band.  Our measurements tend to be systematically higher ($\sim 10\%$) than
the catalog values in \w1, marginally higher (5\%) in \w2, and consistent in \w3
and \w4.  Only the normal, detected objects are included in the statistics.}
\end{center}
\label{fig:cmpaw}
\end{figure*}

For \wise, we compare our measurements with those from the AllWISE catalog,
which are based on PSF profile fitting.  The two sets of measurements should
agree well for point source targets.  As shown in Figure \ref{fig:cmpaw}, our
measurements become gradually more consistent with the catalog results toward
the longer wavelength bands, consistent with the fact that the targets are
marginally resolved in the \w1 and \w2 bands, but increasingly pointlike
in the \w3 and \w4 bands.

\subsection{Herschel}
\label{apd:herschel}

In order to test the robustness of our data reduction and aperture photometry,
we compare our measurements with those listed in the
\herschel/PACS Point Source Catalog \citep[HPPSC;][]{Marton2017arXiv}.  The
HPPSC measurements are based on aperture photometry on the JScanam maps from
the Standard Product Generation (SPG) pipeline.  The catalog provides source
flux densities extracted with aperture radii 6\arcsec\ for 70 and 100 \micron\
and 12\arcsec\ for 160 \micron.  We compare with catalog measurements for which
the S/N was determined from the ``background RMS," which closely follows our
own methodology.  Except for objects we identified as contaminated, 73, 77,
and 59 objects in our QSO2 sample have measurements at 70, 100, and 160
\micron\ listed in IRSA.

\begin{figure*}
\begin{center}
\includegraphics[height=0.25\textheight]{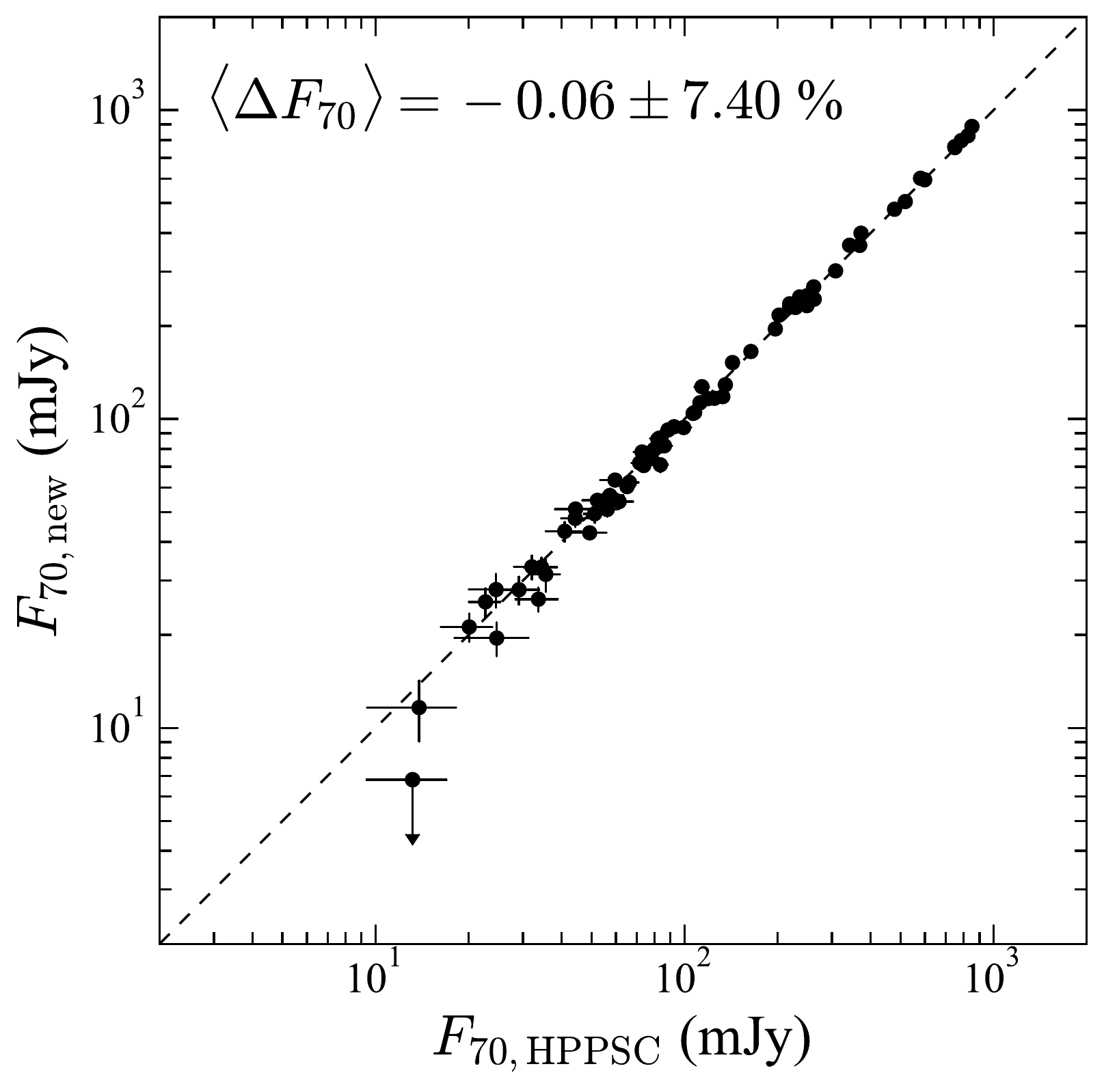}
\includegraphics[height=0.25\textheight]{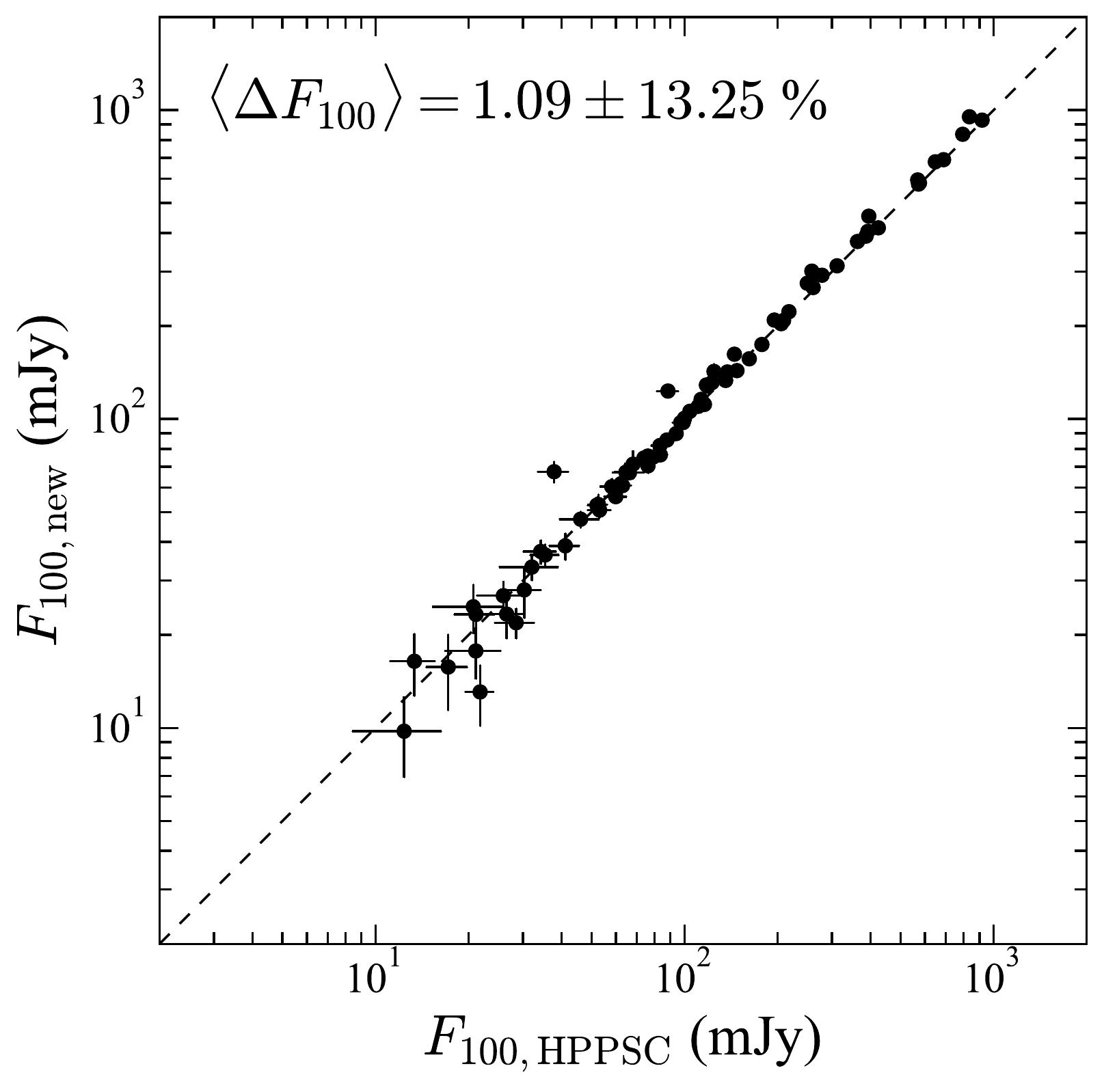}
\includegraphics[height=0.25\textheight]{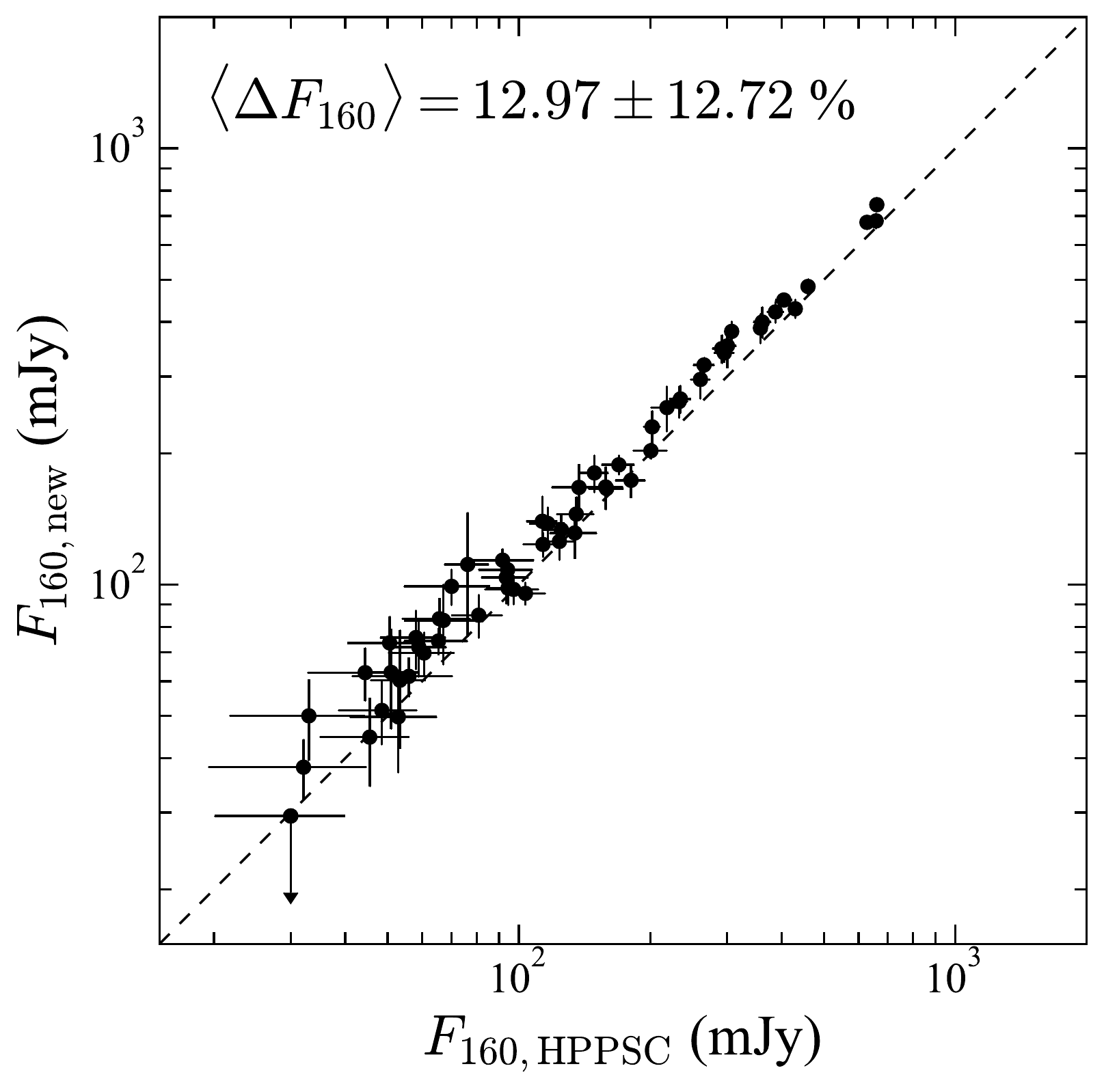}
\caption{Comparison between our new PACS flux densities and those from HPPSC
in the 70, 100, and 160 \micron\ bands.  Objects with flux densities lower
than 3 times of the background RMS are replaced with 3 $\sigma$ as upper
limits.  The median fractional deviation and its standard deviation,
$\langle \Delta F \rangle$, are shown in the upper-left corner of each plot.
The scatter in all three bands is $\sim 10$\%.  For 70 and 100 \micron, the
systematic deviation is negligible.  The $\sim 10$\% systematic deviation at
160 \micron\ is significant; it can be explained by the usage of different
aperture radii and algorithms to generate maps (see main text).}
\end{center}
\label{fig:hppsc}
\end{figure*}

Figure \ref{fig:hppsc} shows that the consistency is very good at 70 and 100
\micron.  The systematic deviations, $\lesssim 1$\%, are much smaller than
the scatter ($\sim 10$\%).  However, at 160 \micron\ the systematic deviation
is comparable to the $\sim 10$\% scatter.  This is due, on the one hand, to
the different aperture sizes we adopt, and, on the other hand, to the different
algorithms used to generate the maps.  If we choose the same aperture set
as HPPSC, the systematic deviation drops to 6.5\%, which can be explained
fully by the differences of the maps generated by JScanam and HPF
methods\footnote{We tested the aperture photometry performed based on three
different algorithms (HPF, JScanam, and Unimap) for generating {\tt level2.5}
maps from the SPG14 pipeline.
We find that the HPF method produces maps with the smallest uncertainty,
although different settings (e.g., the HPF radius) can change the noise level
of the HPF maps.  Maps created using the JScanam and Unimap algorithms show
no systematic deviations relative to HPF at 70 and 100 \micron, but their flux
densities are systematically lower by 6.7\% and 9.1\% at 160 \micron.
Therefore, we adopt the HPF method to generate the maps ourselves, in order to
reduce the noise, but we restrict to the parameters provided by
\cite{Balog2014ExA}.}.  The median uncertainties of the HPPSC measurements are
5.55, 5.39, and 11.92 mJy at 70, 100, and 160 \micron, respectively, while the
corresponding values for our measurements are 3.36, 3.99, and 12.15 mJy.  The
differences can also be explained by the different aperture sizes and the
different algorithms used to generate the maps. 
Our measured uncertainties are also comparable to those reported for the
PG sample of QSO1s \citep{Shangguan2018ApJ}.

\begin{figure*}
\begin{center}
\includegraphics[height=0.25\textheight]{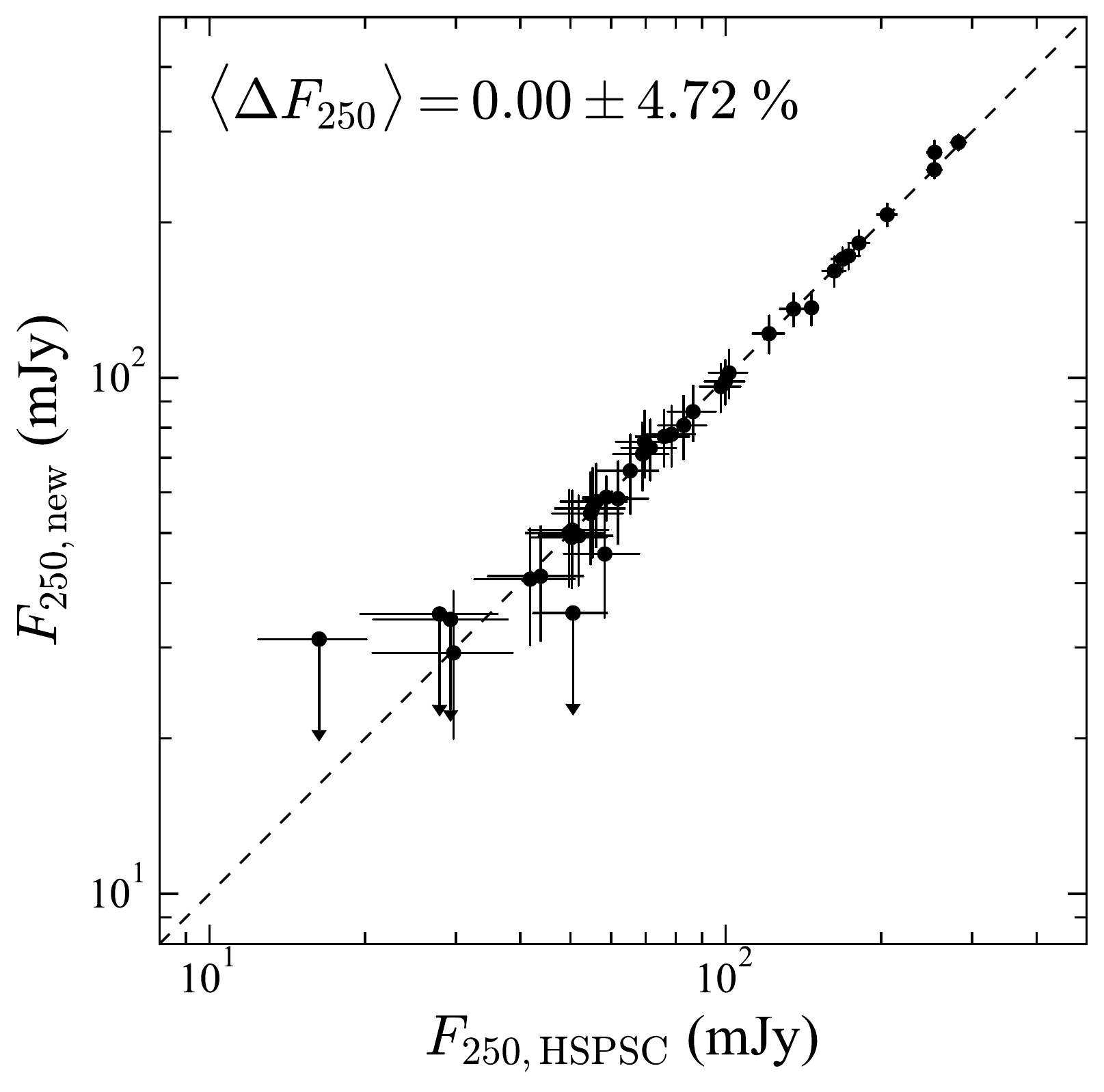}
\includegraphics[height=0.25\textheight]{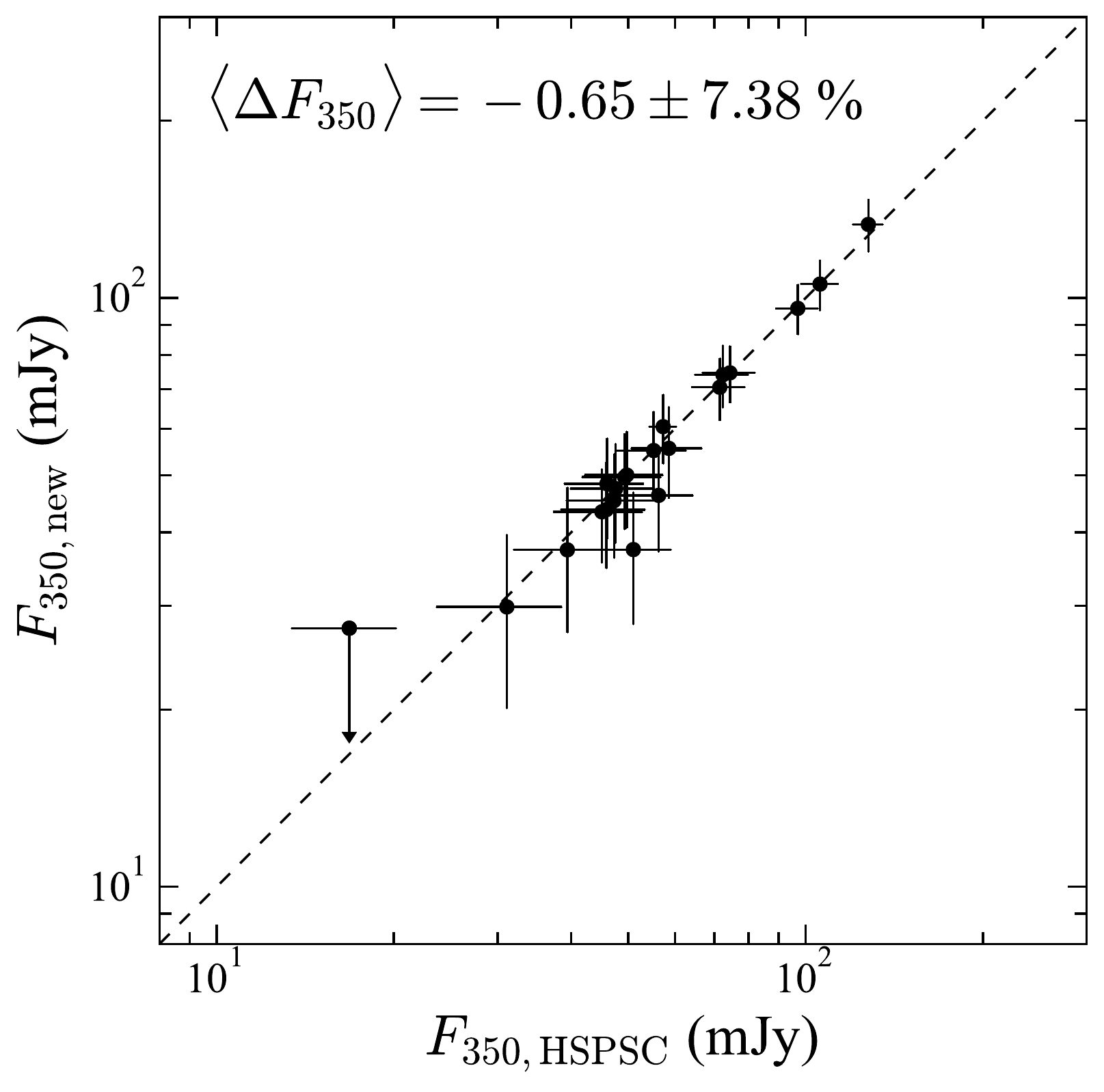}
\includegraphics[height=0.25\textheight]{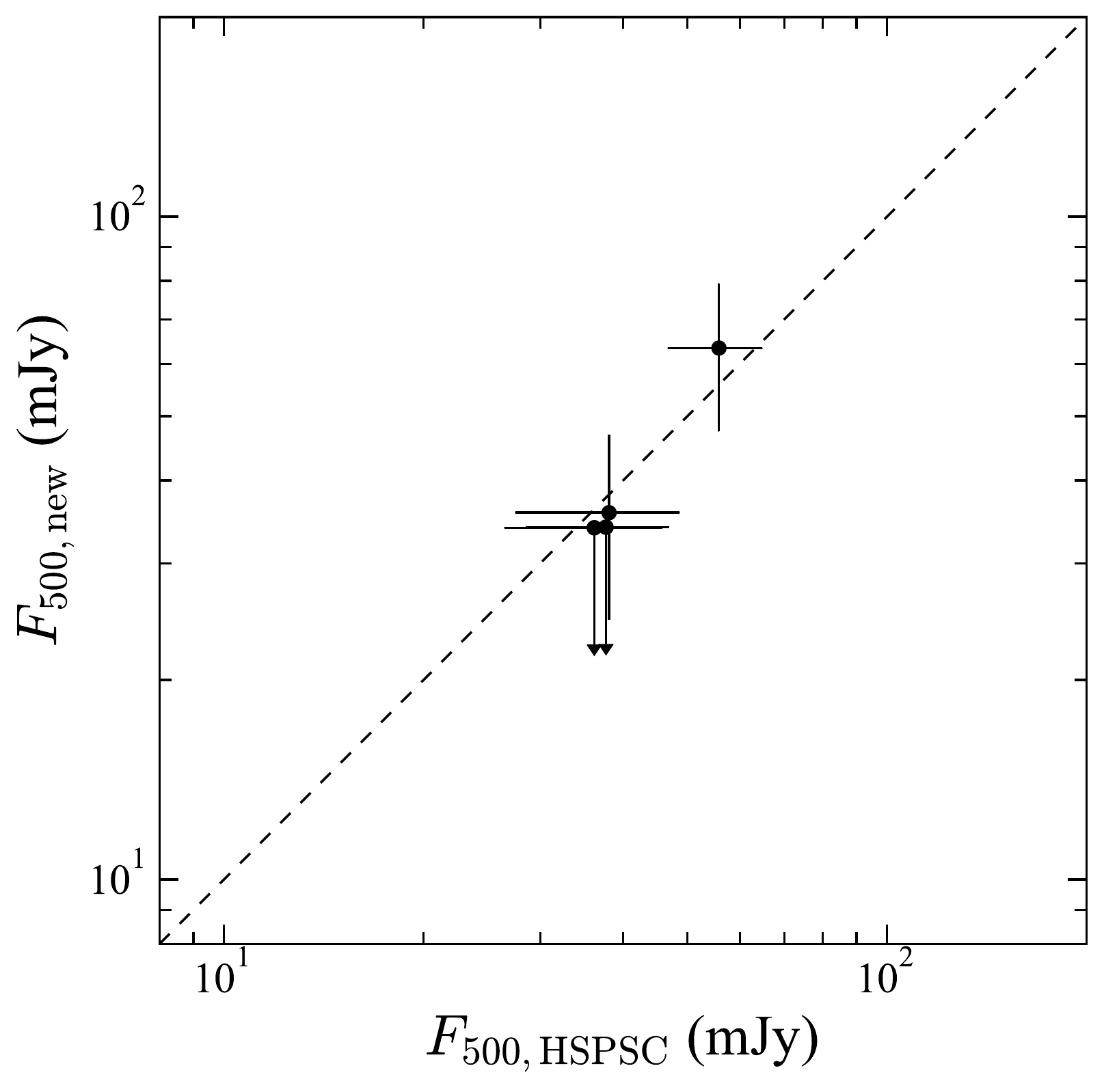}
\caption{Comparison between our new SPIRE flux densities and those from HPPSC
in the 250, 350, and 500 \micron\ bands.  Objects with flux densities lower
than 3 times of the background RMS are replaced with 3 $\sigma$ as upper
limits.  The median fractional deviation and its standard deviation,
$\langle \Delta F \rangle$, are shown in the upper-left corner of each plot.
Only a few objects are undetected; these are not included in the statistics.
For 250 and 350 \micron, the scatter is $\lesssim 10\%$, compared to which the
systematic deviations are negligible.  The number of objects at 500 \micron\
is not enough to draw firm conclusions, although the consistency for the four
objects is still good.}
\end{center}
\label{fig:hspsc}
\end{figure*}

As for SPIRE, 39, 21, and 4 objects at 250, 350, and 500 \micron\ match the
HSPSC within a search radius of 20\arcsec\ from the optical positions of the
QSO2s.  Our measurements in the first two bands are very well
consistent with those in HSPSC (Figure \ref{fig:hspsc}).  For the 500 \micron\
band, there are not enough objects to draw firm conclusions, but the
consistency is still encouraging.

\section{The Influence of Different Dust Torus Models}
\label{apd:torus}

\begin{figure*}
\begin{center}
\includegraphics[height=0.25\textheight]{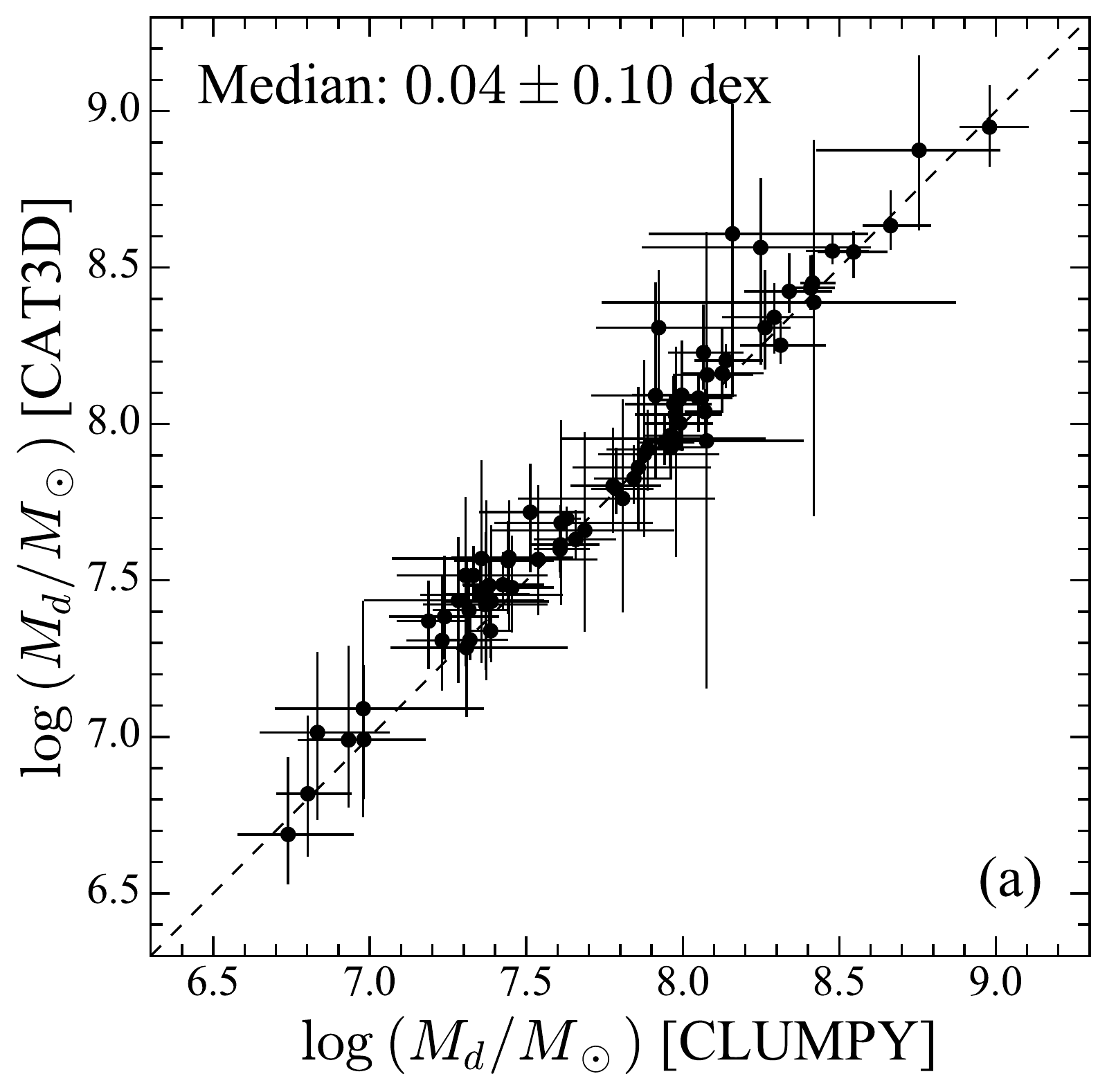}
\includegraphics[height=0.25\textheight]{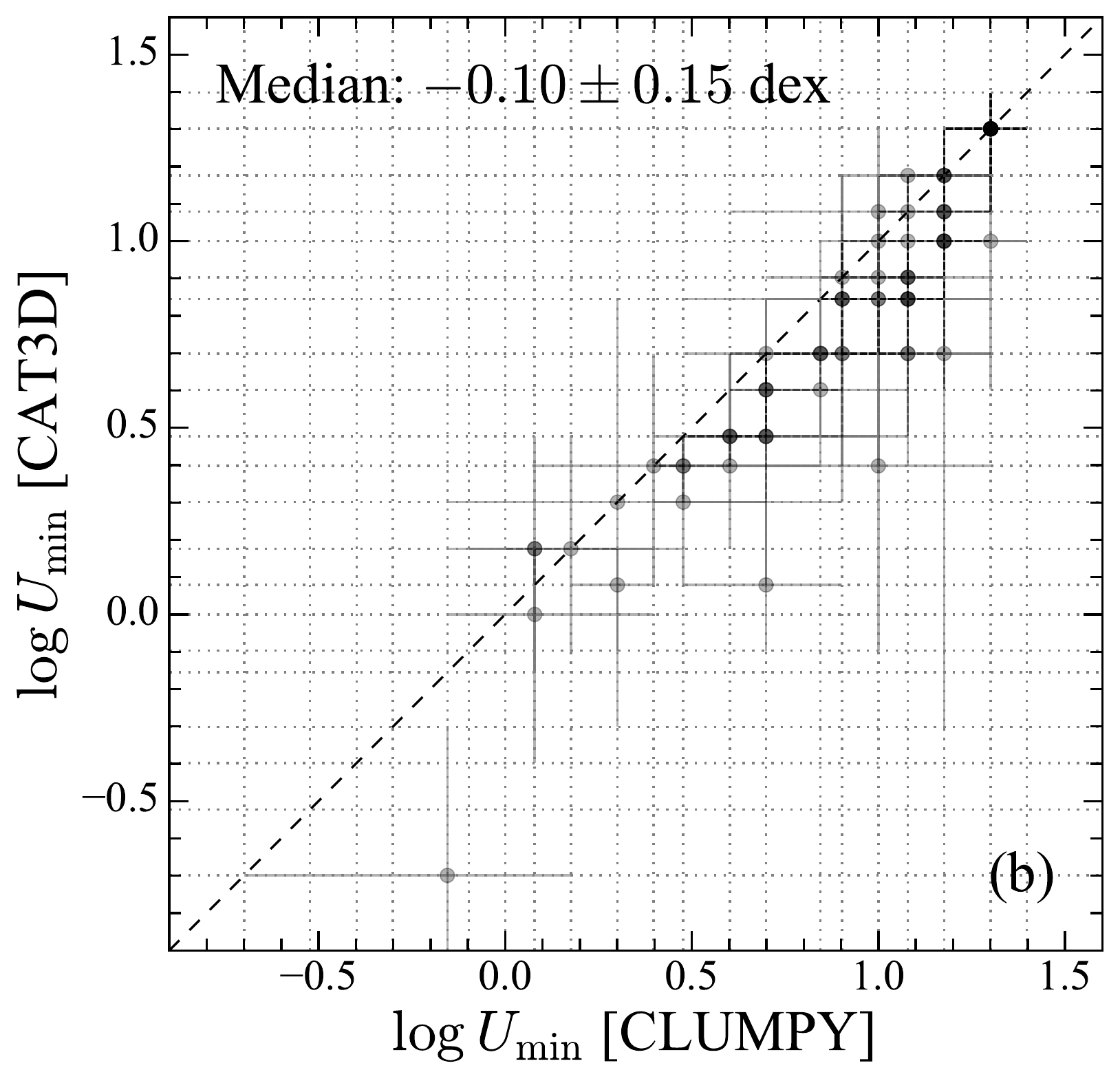}
\includegraphics[height=0.25\textheight]{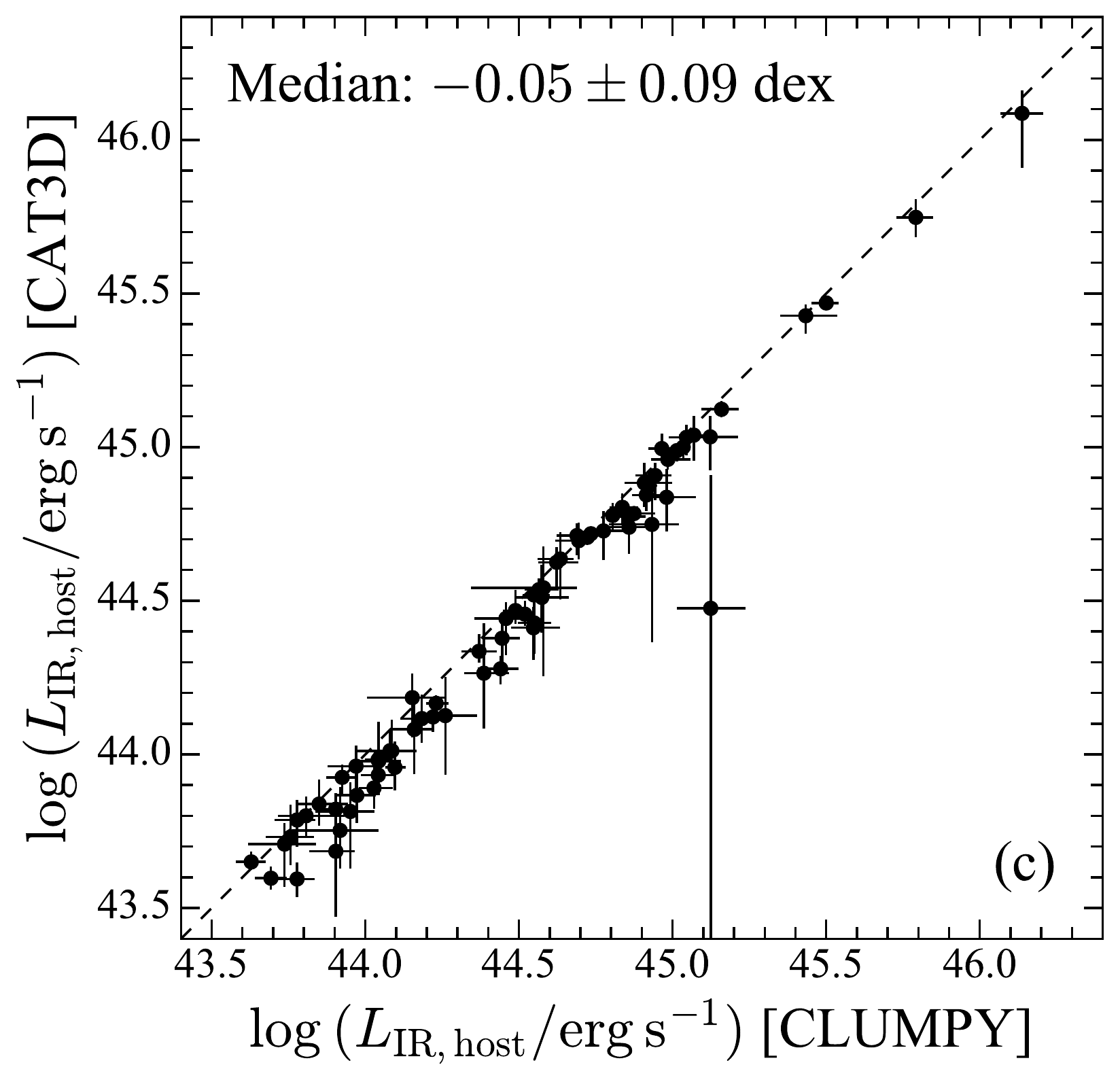}
\caption{SED fits that adopt CAT3D or CLUMPY torus models produce consistent
values of (a) dust mass, (b) $U_\mathrm{min}$, and (c) IR
luminosity of the host galaxy.  The discreteness of $U_\mathrm{min}$ produces
some overlapping data points on the grids; the darker symbols
reflect grids with more data.}
\end{center}
\label{fig:cmpdl07}
\end{figure*}

We evaluate the impact of the choice of dust torus model (CLUMPY vs. CAT3D)
on the derived dust mass, $U_\mathrm{min}$, and IR luminosity.  The CLUMPY
template used in the fits is the median torus template derived by
\PaperI\ based on their SED decomposition of the PG quasars.  We
make this simplifying assumption in order to reduce the number of free
parameters in the fitting.  In general, the CLUMPY template provides more
emission in the \w3 band but less emission in the \w4 band compared to the
data.  As in \PaperI, we add a blackbody component to account for
the possible presence of hot dust emission, which is often needed for QSO1s,
but we find that it is usually negligible for QSO2s.  The CAT3D models of
both \cite{Honig2017ApJ} and \cite{Gonzalez2017MNRAS} generally produce results
consistent with those based on CLUMPY.  In detail, the templates of H\"{o}nig
\& Kishimoto tend to fit the overall SED better, while the templates of
Garc{\'{\i}}a-Gonz{\'a}lez et al. produce fitted parameters that show closer
agreement with those based on CLUMPY.   Figure \ref{fig:cmpdl07} directly
compares the fitting results for the templates of CLUMPY and H\"{o}nig \&
Kishimoto.  The dust masses derived using CAT3D tend to be slightly higher
than those using CLUMPY, while $U_\mathrm{min}$ and $L_\mathrm{IR, host}$
are slightly lower.  This is consistent with the fact that the CLUMPY template
drops quickly toward the FIR, while the best-fit templates from CAT3D are more
extended.  In any event, the systematic differences are quite small, and we
conclude that the key quantities in our work ($M_d$, $U_\mathrm{min}$, and
$L_\mathrm{IR, host}$) are not sensitive to the choice of the dust torus
model.

\clearpage

\startlongtable
\begin{deluxetable}{l r r c c r r r@{$\pm$}l c r@{$\pm$}l}
\tablecaption{Physical Properties of Type 2 Quasars \label{tab:results}}
\tabletypesize{\scriptsize}
\tablehead{
\colhead{Object} &
\colhead{R.A.} &
\colhead{Dec.} &
\colhead{$z$} &
\colhead{$\mathrm{log}\,L_\mathrm{[O\,III]}$} &
\colhead{$\mathrm{log}\,U_\mathrm{min}$} &
\colhead{$\mathrm{log}\,M_d$} &
\mcl{2}{c}{$\mathrm{log}\,M_*$} &
\colhead{$\mathrm{log}\,\delta_\mathrm{GDR}$} &
\mcl{2}{c}{$\mathrm{log}\,M_\mathrm{gas}$} \\
\colhead{} &
\colhead{(J2000)} &
\colhead{(J2000)} &
\colhead{}&
\colhead{($L_\odot$)} &
\colhead{} &
\colhead{($M_\odot$)} &
\mcl{2}{c}{($M_\odot$)} &
\colhead{} &
\mcl{2}{c}{($M_\odot$)} \\
\colhead{(1)} &
\colhead{(2)} &
\colhead{(3)} &
\colhead{(4)}&
\colhead{(5)} &
\colhead{(6)} &
\colhead{(7)} &
\mcl{2}{c}{(8)} &
\colhead{(9)} &
\mcl{2}{c}{(10)}
}
\startdata
 SDSS J0041$-$0952  & 00 41 13.75   & $-$09 52 31.7   &  0.095  &      8.31   & $   0.40_{-0.22}^{+0.20}$  &   $7.53_{-0.21}^{+0.18}$  &     10.89 &    0.22  &  2.09  &     9.61  &     0.28 \\
 SDSS J0119$-$1026  & 01 19 35.64   & $-$10 26 13.1   &  0.125  &      8.40   & $   0.70_{-0.22}^{+0.30}$  &   $7.57_{-0.24}^{+0.19}$  &     10.67 &    0.22  &  2.09  &     9.65  &     0.30 \\
 SDSS J0202$+$1247  & 02 02 23.71   & $+$12 47 17.7   &  0.086  &      8.11   & $   0.48_{-0.08}^{+0.22}$  &   $8.09_{-0.12}^{+0.09}$  &     10.77 &    0.24  &  2.09  &    10.18  &     0.22 \\
 SDSS J0339$-$0548  & 03 39 23.15   & $-$05 48 41.6   &  0.085  &      8.22   & $   0.84_{-0.15}^{+0.06}$  &   $7.34_{-0.06}^{+0.14}$  &     10.30 &    0.22  &  2.09  &     9.43  &     0.22 \\
 SDSS J0747$+$3200  & 07 47 51.57   & $+$32 00 52.1   &  0.280  &      8.93   & $   0.84_{-0.15}^{+0.06}$  &   $8.43_{-0.08}^{+0.14}$  &     11.12 &    0.25  &  2.09  &    10.52  &     0.23 \\
 SDSS J0752$+$2510  & 07 52 26.52   & $+$25 10 20.2   &  0.239  &      8.69   & $   1.00_{-0.30}^{+0.18}$  &   $8.07_{-0.17}^{+0.19}$  &     11.15 &    0.25  &  2.09  &    10.16  &     0.27 \\
 SDSS J0753$+$3847  & 07 53 24.38   & $+$38 47 31.8   &  0.097  &      8.21   & $   0.70_{-0.10}^{+0.15}$  &   $7.81_{-0.13}^{+0.11}$  &     10.93 &    0.23  &  2.09  &     9.90  &     0.23 \\
 SDSS J0753$+$2309  & 07 53 29.94   & $+$23 09 30.7   &  0.337  &      8.57   &     \mcl{1}{c}{\nodata}    &                $<8.70$    & \mcl{2}{r}{$<$11.23} &  2.08  & \mcl{2}{r}{$<$10.78} \\
\enddata
\tablecomments{
(1) Source name.
(2) and (3) J2000 coordinates.
(4) Redshift.
(5) \OIIIb\ luminosity from \cite{Reyes2008AJ}.
(6) Best-fit minimum intensity of the interstellar radiation field relative to that measured in
the solar neighborhood.
The quoted uncertainties represent the 68 percent confidence interval determined from the 16th and
84th percentile of the marginalized posterior probability density function.  However, if there are fewer 
than 16\% sampled values at the discrete grids below (above) the best-fit value, the lower (upper) 
uncertainty of the parameter is not resolved, and it is reported as ``0.00'' in the table.
(7) Best-fit total dust mass.
(8) Stellar mass derived from optical color and $J$-band absolute magnitude \citep{BdJ01}, converted to \cite{Chabrier2003PASP} IMF.
(9) Gas-to-dust ratio estimated in Section \ref{ssec:imass}.
(10) Total gas mass including helium and heavier elements.
} 
\end{deluxetable}


\startlongtable
\begin{deluxetable*}{l c r@{$\pm$}l r@{$\pm$}l r@{$\pm$}l r@{$\pm$}l r@{$\pm$}l r@{$\pm$}l r@{$\pm$}l}
\tabletypesize{\scriptsize}
\tablecaption{NIR and MIR Photometry \label{tab:nirdata}}
\tablehead{
\colhead{Object} &
\colhead{$r_\mathrm{2MASS}$} &
\mcl{2}{c}{$F_{J}$} &
\mcl{2}{c}{$F_{H}$} &
\mcl{2}{c}{$F_{K_s}$} &
\mcl{2}{c}{$F_{W1}$} &
\mcl{2}{c}{$F_{W2}$} &
\mcl{2}{c}{$F_{W3}$} &
\mcl{2}{c}{$F_{W4}$} \\
\colhead{} &
\colhead{(\arcsec)} &
\mcl{2}{c}{(mJy)} &
\mcl{2}{c}{(mJy)} &
\mcl{2}{c}{(mJy)} &
\mcl{2}{c}{(mJy)} &
\mcl{2}{c}{(mJy)} &
\mcl{2}{c}{(mJy)} &
\mcl{2}{c}{(mJy)} \\
\colhead{(1)} &
\colhead{(2)} &
\mcl{2}{c}{(3)} &
\mcl{2}{c}{(4)} &
\mcl{2}{c}{(5)} &
\mcl{2}{c}{(6)} &
\mcl{2}{c}{(7)} &
\mcl{2}{c}{(8)} &
\mcl{2}{c}{(9)}
}
\startdata
SDSS J0041$-$0952 &  20 &   3.02 &      0.53\ph{0}  &   3.97 &      0.85\ph{0}  &   5.94 &      1.01\ph{0}   &   1.91 &      0.01\ph{0}  &   1.29 &      0.02\ph{0}  &   2.24 &      0.15\ph{00} &  14.87 &      1.46\ph{00}  \\ 
SDSS J0119$-$1026 &   7 &   1.00 &      0.18        &   1.45 &      0.26        &   1.65 &      0.34         &   1.02 &      0.01        &   1.08 &      0.03        &   4.93 &      0.14        &  20.61 &      1.23  \\ 
SDSS J0202$+$1247 &  20 &   2.96 &      0.77        &   4.26 &      1.15        & \mcl{2}{c}{$<$4.11}        &   4.52 &      0.02        &   6.20 &      0.02        &  22.73 &      0.13        &  75.14 &      1.06  \\ 
SDSS J0339$-$0548 &  10 &   1.00 &      0.19        &   1.17 &      0.28        &   1.51 &      0.42         &   0.93 &      0.01        &   0.98 &      0.02        &   6.23 &      0.15        &  21.08 &      1.16  \\ 
SDSS J0747$+$3200 &   7 &   0.42 &      0.13        &   0.90 &      0.19        &   0.86 &      0.26         &   0.42 &      0.02        &   0.44 &      0.02        &   3.47 &      0.16        &  17.97 &      1.47  \\ 
SDSS J0752$+$2510 &   7 &   0.69 &      0.19        & \mcl{2}{c}{$<$0.86}       & \mcl{2}{c}{$<$0.89}        &   0.97 &      0.02        &   1.68 &      0.03        &   8.14 &      0.18        &  29.53 &      1.65  \\ 
SDSS J0753$+$3847 &  20 &   3.15 &      0.69        &   3.53 &      1.04        & \mcl{2}{c}{$<$3.29}        &   1.51 &      0.02        &   1.30 &      0.03        &   4.98 &      0.13\tnm{a} &  20.91 &      1.71  \\ 
SDSS J0753$+$2309 &   7 & \mcl{2}{c}{$<$0.33}       & \mcl{2}{c}{$<$0.59}       & \mcl{2}{c}{$<$0.65}        &   0.21 &      0.02        &   0.24 &      0.02        &   1.33 &      0.17        &  10.16 &      1.38  \\
\enddata
\tablenotetext{a}{The flux density is measured after \galfit\ decomposition.}
\tablecomments{
(1) Source name.
(2) The aperture radius of \tmass\ aperture photometry.
(3) The flux density and $1\,\sigma$ uncertainty at $J$ band.
(4) The flux density and $1\,\sigma$ uncertainty at $H$ band.
(5) The flux density and $1\,\sigma$ uncertainty at $K_s$ band.
(6) The flux density and $1\,\sigma$ uncertainty at $W1$ band.
(7) The flux density and $1\,\sigma$ uncertainty at $W2$ band.
(8) The flux density and $1\,\sigma$ uncertainty at $W3$ band.
(9) The flux density and $1\,\sigma$ uncertainty at $W4$ band.
The objects fainter than $3\,\sigma$ of the sky variation are replaced by the
$3\,\sigma$ as upper limits.
}
\end{deluxetable*}

\startlongtable
\begin{deluxetable*}{l c r@{$\pm$}l r@{$\pm$}l r@{$\pm$}l r@{$\pm$}l r@{$\pm$}l r@{$\pm$}l}
\tabletypesize{\scriptsize}
\tablecaption{{\it Herchel} Photometry of Type 2 Quasars \label{tab:herschel}}
\tablehead{
\colhead{Object} &
\colhead{Aperture} &
\mcl{2}{c}{$F_{70}$} &
\mcl{2}{c}{$F_{100}$} &
\mcl{2}{c}{$F_{160}$} &
\mcl{2}{c}{$F_{250}$} &
\mcl{2}{c}{$F_{350}$} &
\mcl{2}{c}{$F_{500}$} \\
\colhead{} &
\colhead{} &
\mcl{2}{c}{(mJy)} &
\mcl{2}{c}{(mJy)} &
\mcl{2}{c}{(mJy)} &
\mcl{2}{c}{(mJy)} &
\mcl{2}{c}{(mJy)} &
\mcl{2}{c}{(mJy)} \\
\colhead{(1)} &
\colhead{(2)} &
\mcl{2}{c}{(3)} &
\mcl{2}{c}{(4)} &
\mcl{2}{c}{(5)} &
\mcl{2}{c}{(6)} &
\mcl{2}{c}{(7)} &
\mcl{2}{c}{(8)}
}
\startdata
 SDSS J0041$-$0952        &   l &     31.40  &      3.90\ph{00} &      67.53  &       5.44        &      62.87  &    16.23\ph{0}  &   29.31  &   9.39\ph{0}   &  \mcl{2}{c}{$<$24.60}     &       \mcl{2}{c}{$<$33.61}       \\
 SDSS J0119$-$1026        &   s &     54.04  &      3.17        &      61.76  &       3.66        &      74.22  &     5.31        &  \mcl{2}{c}{$<$32.98}     &  \mcl{2}{c}{$<$25.98}     &       \mcl{2}{c}{$<$31.67}       \\
 SDSS J0202$+$1247        &   l &    236.06  &      4.33        &     301.05  &       7.06\ph{00} &     347.36  &    26.11        &  161.56  &  10.57         &   88.33  &   8.94         & \ph{$^a$}34.01  &  10.71           \\
 SDSS J0339$-$0548        &   s &     85.99  &      2.59        &     111.50  &       3.52        &     108.10  &     8.11        &   57.71  &  10.46         &  \mcl{2}{c}{$<$24.99}     &       \mcl{2}{c}{$<$31.78}       \\
 SDSS J0747$+$3200        &   s &     74.81  &      2.24        &     115.78  &       3.06        &     123.59  &     8.14        &   73.17  &  10.07         &   29.74  &   9.15         &       \mcl{2}{c}{$<$33.15}       \\
 SDSS J0752$+$2510        &   s &     81.89  &      2.71        &      89.71  &       3.23        &      99.05  &     9.55        &   46.85  &  11.49         &  \mcl{2}{c}{$<$29.44}     &       \mcl{2}{c}{$<$45.48}       \\
 SDSS J0753$+$3847\tnm{a} &   e &    127.22  &      5.32        &     158.53  &      10.07        &     202.09  &    34.62        &   80.86  &  11.63\tnm{c}  &   55.08  &   9.11\tnm{c}  &       \mcl{2}{c}{$<$33.50}       \\
 SDSS J0753$+$2309        &   s &     26.11  &      2.46        &      21.91  &       2.50        &   \mcl{2}{c}{$<$32.93}        &  \mcl{2}{c}{$<$32.20}     &  \mcl{2}{c}{$<$25.75}     &       \mcl{2}{c}{$<$30.95}       \\
\enddata
\tablenotetext{a}{All the bands are $>5\%$ affected by the contamination.}
\tablenotetext{b}{The 160 \micron\ band is $>5\%$ affected by the contamination.}
\tablenotetext{c}{The measurement may be contaminated. Since the companions 
are always much fainter than the quasar in PACS maps, we believe that the
contamination is not significant.}
\tablecomments{
(1) Source name.
(2) The aperture type adopted to extract the flux. ``s'', ``l'', and ``e'' refer to small, large, and extended apertures, respectively.
(3) The flux density and $1\,\sigma$ uncertainty at 70 \micron.
(4) The flux density and $1\,\sigma$ uncertainty at 100 \micron.
(5) The flux density and $1\,\sigma$ uncertainty at 160 \micron.
(6) The flux density and $1\,\sigma$ uncertainty at 250 \micron.
(7) The flux density and $1\,\sigma$ uncertainty at 350 \micron.
(8) The flux density and $1\,\sigma$ uncertainty at 500 \micron.
The objects fainter than $3\,\sigma$ of the sky variation are replaced by the
$3\,\sigma$ as upper limits.
}
\end{deluxetable*}

\begin{table*}
  \begin{center}
  \caption{Model Parameters and Priors}
  \label{tab:pars}
  \begin{tabular}{cccclp{9cm}}
    \hline
    \hline
    Models                       & Parameters       & Units                  & Discreteness & \mcl{1}{c}{Priors}         & \mcl{1}{c}{Descriptions}                                                                \\
    (1)                          & (2)              & (3)                    & (4)          & \mcl{1}{c}{(5)}            & \mcl{1}{c}{(6)}                                                                         \\\hline
    \multirow{2}{*}{BC03}        & $M_*$            & $M_\odot$              & \ding{56}    & [$10^6$, $10^{14}$]        & Stellar mass.                                                                           \\
                                 & $t$              & Gyr                    & \ding{52}    & 5 (fixed)                  & The age of the single stellar population.                                                      \\\hline
    \multirow{5}{*}{CAT3D}       & $i$              & --                     & \ding{52}    & [0.0, 90.0]                & The inclination angle of the torus.                                                     \\
                                 & $a$              & --                     & \ding{52}    & [-2.5, -0.25]              & The power-law index of the dust cloud radial distribution.                              \\
                                 & $N_0$            & --                     & \ding{52}    & [5.0, 10.0]                & The number of dust clouds along an equatorial line-of-sight.                            \\
                                 & $h$              & --                     & \ding{52}    & [0.25, 1.50]               & The dimensionless scale height of the vertical Gaussian distribution of the dust cloud. \\
\hline
  \end{tabular}
  \end{center}
\tablecomments{(1) The name of the models used in the SED fitting.
(2) The parameters of each model.
(3) The units of the parameters.
(4) Whether the parameter is discrete and requires interpolation to implement the MCMC fitting.
(5) The prior range of the parameters.
(6) A brief description of the parameters.}
\end{table*}

\begin{table*}
  \begin{center}
  \caption{Statistics of dust, gas, and stellar masses}
  \label{tab:med}
  \begin{tabular}{cccc}
    \hline
    \hline
    \mcl{4}{c}{Kaplan-Meier estimator}                                        \\\hline
                  & $\log\,M_d$            & $\log\,M_\mathrm{gas}$ & $\log\,M_*$             \\
                  & ($M_\odot$)            & ($M_\odot$)            & ($M_\odot$)             \\\hline
    Type 1 quasar & $7.72_{-0.31}^{+0.32}$ & $9.82_{-0.32}^{+0.32}$ & $10.90_{-0.18}^{+0.24}$ \\
    Type 2 quasar & $7.74_{-0.28}^{+0.36}$ & $9.82_{-0.29}^{+0.36}$ & $10.93_{-0.25}^{+0.19}$ \\\hline\hline
    \mcl{4}{c}{Monte Carlo method}                                                            \\\hline
                  & $\log\,M_d$            & $\log\,M_\mathrm{gas}$ & $\log\,M_*$             \\
                  & ($M_\odot$)            & ($M_\odot$)            & ($M_\odot$)             \\\hline
    Type 1 quasar & $7.69_{-0.04}^{+0.04}$ & $9.78_{-0.05}^{+0.07}$ & $10.92_{-0.05}^{+0.04}$ \\
    Type 2 quasar & $7.76_{-0.04}^{+0.04}$ & $9.83_{-0.05}^{+0.06}$ & $10.88_{-0.05}^{+0.04}$ \\\hline
  \end{tabular}
  \end{center}
\tablecomments{
Upper part: The $50_{-25}^{+25}$th percentiles of the dust, gas, and 
stellar masses are calculated with the Kaplan-Meier estimator.  
Lower part: The $50_{-25}^{+25}$th percentiles of the median values 
are calculated using a Monte Carlo method to resample the data based 
on the measurement and uncertainty of each object, assuming a 
Gaussian distribution; the median of each quantity is obtained using 
the Kaplan-Meier estimator.}
\end{table*}

\end{document}